\documentclass[pra,twocolumn,a4paper,showkeys]{revtex4-1}

\usepackage{latexsym}
\usepackage{amssymb}
\usepackage{amsmath}
\usepackage{enumerate}
\usepackage{bm}
\usepackage{paralist}

\usepackage[latin1]{inputenc}
\usepackage[normalem]{ulem}

\usepackage{xcolor}
\usepackage{subfigure}
\usepackage {graphicx}
\graphicspath{{./figs/}}

\newcommand {\qfigs}[1]{Figs.~\ref{#1}}
\newcommand {\qfig}[1]{Fig.~\ref{#1}}
\newcommand {\qsect}[1]{Sect.~\ref{#1}}

\newcommand {\beql}[1]{\begin{equation} \label{#1}}
\newcommand {\eeql}{\end{equation}}
\newcommand {\beq}{\begin{equation}}
\newcommand {\eeq}{\end{equation}}

\newcommand {\etal}{\emph{et al.}\ }

\begin{document}

\date{\today}
\title{Morphology of graphene flakes in Ni-graphene nanocomposites and its influence on hardness: an atomistic study}

\author{Vardan Hoviki Vardanyan}
\affiliation{%
Physics Department,
University Kaiserslautern,
Erwin-Schr{\"o}dinger-Stra{\ss}e, D-67663 Kaiserslautern, Germany}
\affiliation{%
Research Center OPTIMAS,
University Kaiserslautern,
Erwin-Schr{\"o}dinger-Stra{\ss}e, D-67663 Kaiserslautern, Germany}

\author{Herbert M.~Urbassek}
\email{urbassek@rhrk.uni-kl.de}
\homepage{http://www.physik.uni-kl.de/urbassek/}
\affiliation{%
Physics Department,
University Kaiserslautern,
Erwin-Schr{\"o}dinger-Stra{\ss}e, D-67663 Kaiserslautern, Germany}
\affiliation{%
Research Center OPTIMAS,
University Kaiserslautern,
Erwin-Schr{\"o}dinger-Stra{\ss}e, D-67663 Kaiserslautern, Germany}

\begin{abstract}

The effect of graphene flakes on the strength of Ni-graphene composites is investigated using molecular dynamics simulation.
Rather than introducing flakes as flat structures into the Ni matrix, as it is common in available studies, we introduce them into the heated liquid Ni and let the structures equilibrate in a 14-ns  molecular-dynamics run; these structures are then quenched to obtain the composites. By varying the interaction of flake edge atoms with the Ni matrix, two different flake morphologies -- wrinkled vs flat -- are obtained. The mechanical properties, and  in particular the composite hardness, are investigated by a simulated nanoindentation test. The flake morphology affects the plastic activity and the hardness of the composites. Wrinkled flakes show a higher potential in absorbing dislocations  than flat flakes, resulting in a considerably reduced hardness of the composite. No effect of the changed interaction between graphene edge atoms and the Ni matrix on the dislocation activity and hardness could be observed. Furthermore, we demonstrate that a high graphene content in the plastic zone leads to an increased dislocation absorption and weakens the composite.

\end{abstract}


\maketitle


\section{Introduction}

Graphene-metal composites have excellent mechanical properties  \cite{XCG*15,RAS*08,ZMF*14,YWLH16,GKH19,FSY*17,LWWW16} as they combine the superior mechanical properties of graphene \cite{LWKH08,CFH*20} with the ductility of the matrix metal. The increasing interest in graphene-based metal nanocomposites \cite{FGL*17,HYJ*13,KTA*15} requires a better understanding  of the hardening mechanisms in such materials. 

The dislocation-interface interaction is key to understanding the   mechanical properties of graphene-metal composites. Experimental studies indicate the importance of dislocation pile-up and blocking at the  metal-graphene interface as the source of the graphene-induced hardening in the composite  \cite{HKK*17,LGL*15,KLY*13}.

Molecular dynamics simulations   have been proven as an excellent method to study the dislocation-interface  interaction, such as for metal-amorphous interfaces \cite{ZC16,SKD*17,SXZ*17,AVKU21},  metallic multilayers \cite{WM11} or  poly-crystalline systems \cite{SM09,STJ98,SW06,ZFZ*20}. Several molecular dynamics studies were devoted to investigating   graphene-metal nanocomposites \cite{CNB13,MSN18,LWWW16,YBM17,VU19,VU20,SA21,SDA21,ZHW20}. The considerable  improvement   of the composite hardness  \cite{PS20,Rez18}  was explained by the hindering of dislocation propagation in the metal matrix by graphene \cite{CNB13,YBM17,MZX*20}. In addition, the importance of the absorption of dislocations by the interface was reported  \cite{SA21,SA20} and it was found  that dislocation absorption can weaken the system \cite{VU19,VU20}.

In further detail,   Shuang and Aifantis \cite{SA21} pointed out that the edges of graphene flakes are important for the composite strength as they act as dislocation sources in compression experiments. They also  argued that the previously reported  improvement  of hardening \cite{ZLP*19,RDTS16,ZLP*19a}  may be  an artifact of the simulation design, if graphene flakes crossing the simulation cell with periodic boundary are used. The authors recommend to use free graphene flakes in future simulations.

Xia \etal \cite{XDZ*21} demonstrated that the strength of graphene-metal nanocomposites depends on the metal grain size and the graphene volume concentration.  Wei \etal \cite{WYH*21} showed that there is an optimum concentration of graphene for better mechanical performance such that any further increase of the graphene content lowers the mechanical resistivity of the  composite.
  
The previously published research  in metal-graphene composites \cite{HSA21,RHZ*18,ZLWY21,WJY*21,ZAL*20,Kum18,SA21,SDA21,ZHW20,ZWH21,VU19,VU20} lacks a discussion of the influence of  the graphene concentration on  the composite  hardness, and in particular on a transition from hardening to weakening of the composite structures. As a rule, if the simulated composite  shows hardening by graphene reinforcement, a further increase of the graphene content further increases the hardness \cite{WNF*18,ZLP*19,MZX*20,SDA21}; and analogous for a weakening effect \cite{CNB13,ZAL*20}. This  contradicts numerous experimental findings \cite{YLZ*13,CJ14,HTL*16,ZZ16,CYW*16,GYG*16,LX17,CZL*16} which indicate  that there exists an optimum graphene fraction in the component to maximize the hardness.

Most previous studies use a  flat morphology of graphene flakes. However, it well known that graphene has a wrinkled structure in composites \cite{GYG*16,SWS*20} and molecular dynamics simulations also show easy wrinkling of graphene \cite{VF21}. The bending stiffness of a single graphene layer was  reported  \cite{WWW*13,DB16,CGY*19, VF21} and it was shown how  the difference between in-plane stiffness and out-of-plane bending stiffness allows    crumpling of graphene flakes   and the formation of  stretchable structures \cite{BBR*15,CFH*20}. Moreover, experimental studies show that    graphene crumpling    has   a significant effect on  softening a  free-standing graphene flake \cite{NCL*15}. Recently,   Kumar \etal \cite{KPD19} studied aggregation of graphene  flakes in a liquid aluminum matrix and showed a way for simulations to obtain more complex flake structures than the usually used flat shape. 

In the present study, we focus on the effect of the morphology of graphene flakes on the composite hardness. By modifying the graphene-Ni interaction, graphene flakes change their morphology  in the liquid Ni melt and keep this morphology during quenching into the solid state. Simulated nanoindentation of these nanocomposite structures shows that the morphology strongly influences the composite hardness. In addition, we  evaluate the effect of the local graphene content in the plastic zone under the indenter, and of pre-existing dislocations on the hardening and weakening of the composite with respect to a polycrystalline Ni sample. 

\section{Method} \label{s_m}

\subsection{Preparation of the systems} 





\begin{figure}[ht]
\begin{center}

\includegraphics[width=0.9\columnwidth]{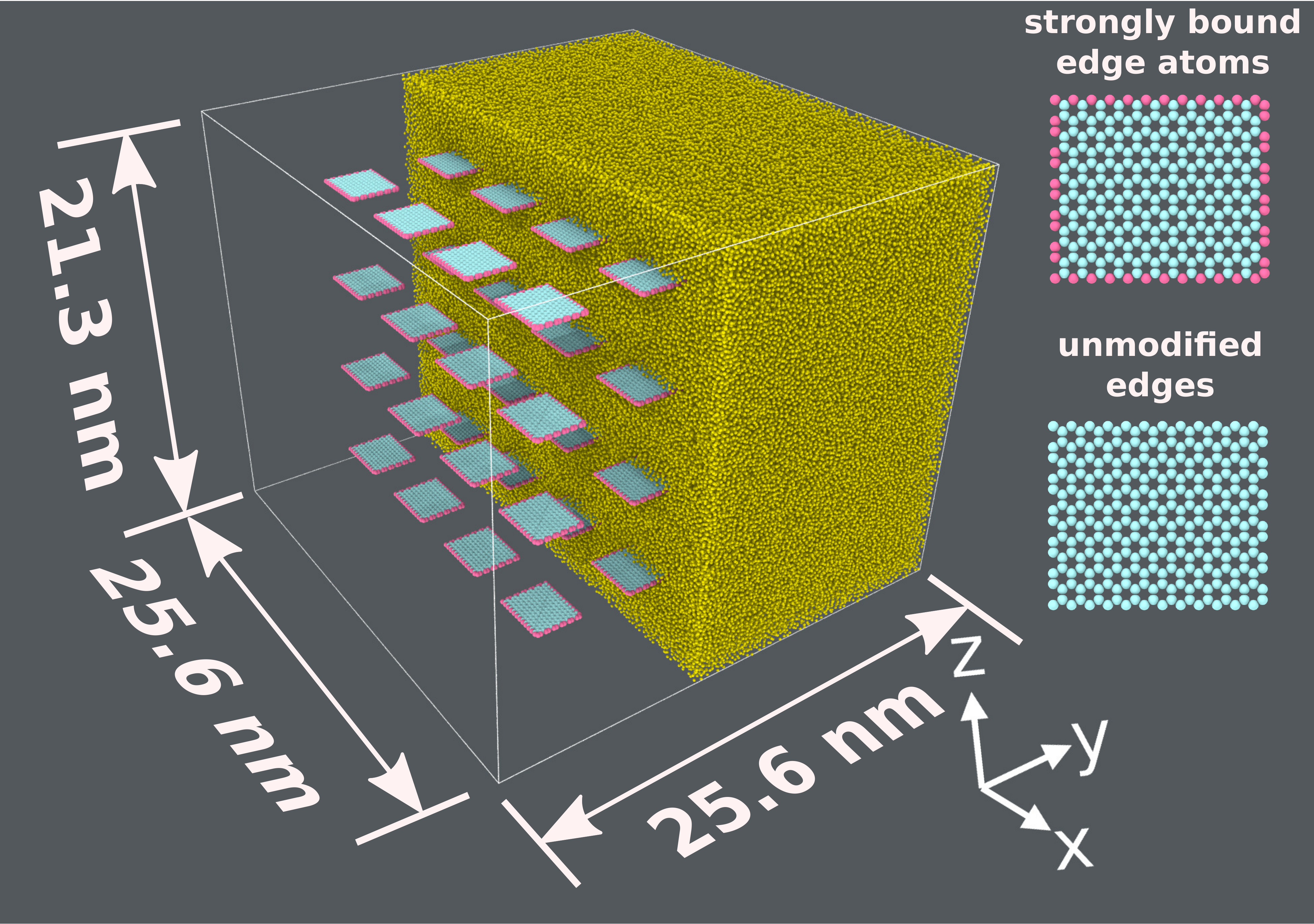}
\end{center}
\caption{Preparation of the Ni-graphene nanocomposite system. The cuboid simulation volume is filled with liquid Ni atoms (yellow) and graphene flakes (blue); graphene edge atoms with a modified C-Ni interaction are colored red.  Half of the nickel atoms are removed for clarity. The inset shows schematically the difference in our models of graphene: at the top graphene with strongly bound edge atoms; at the bottom a graphene flake with unmodified edge atoms
see text. 
}
\label{f_setup}
\end{figure}

The system preparation starts with  a liquid  block of nickel  at  2300 K with dimensions of $25.6 \times 25.6 \times 21.3$ nm, containing  approximately $1.1 \times 10^6$ atoms. 64 Graphene flakes are evenly distributed in the liquid, see \qfig{f_setup}. Each flake has a  square shape  with edge length 26 \AA\ containing 288 carbon atoms. Thus the atomic fraction of graphene in  the composite amounts to $1.6$ \%.

In our simulation the interaction between Ni atoms is modeled by the  interaction potential developed by Mishin \etal  \cite{MFMP99}, and the interaction between carbon atoms by the  bond-order  potential of Stuart \etal \cite{STH00}. The Ni-C interaction is  modeled by a  Lennard-Jones (LJ)  potential. Here, we consider two choices:
\begin{enumerate}[(i)]
\item All C atoms interact via the same LJ potential. We choose the parameters as given by Huang \etal  \cite{HMB03}: $\varepsilon=23.049$ meV, $\sigma = 2.852$ \AA. In the following, we will denote this interaction  and the embedded graphene flakes as \emph{graphene with unmodified edges}.
\item It is known that edge atoms of the graphene flakes can interact covalently with Ni. Tavazza \etal \cite{TSZ*15} calculated the Ni-C interaction by DFT and fitted their results to a LJ potential with the parameters $\varepsilon=200$ meV,  $\sigma = 1.514$ \AA. We will denote graphene flakes as  \emph{graphene with strongly bound edge atoms}, in which edge atoms underlie this increased interaction, while the inner graphene atoms feel the weak LJ potential. 
\end{enumerate}

The difference of our models of weak and strong graphene  is sketched in the inset of  \qfig{f_setup}.

We note that in previous studies of the Ni-graphene system \cite{KPD19,VU19,VU20}, the effects of a changed interaction potential between C and Ni atoms were studied, but it was changed for all C atoms, including those in the interior of the flakes. Here for the first time, we differentiate between interior flake atoms and edge atoms, and investigate to what extent their different interaction with Ni influences the composite preparation and the mechanical properties of the composite.

Besides the Ni-graphene composites, we also prepared a pure Ni system with the same size and following the same preparation procedure for comparison purposes.

\subsection{Equilibration}
 
\begin{figure}[ht]
\begin{center}
\subfigure[]{\includegraphics[width=0.9\columnwidth]{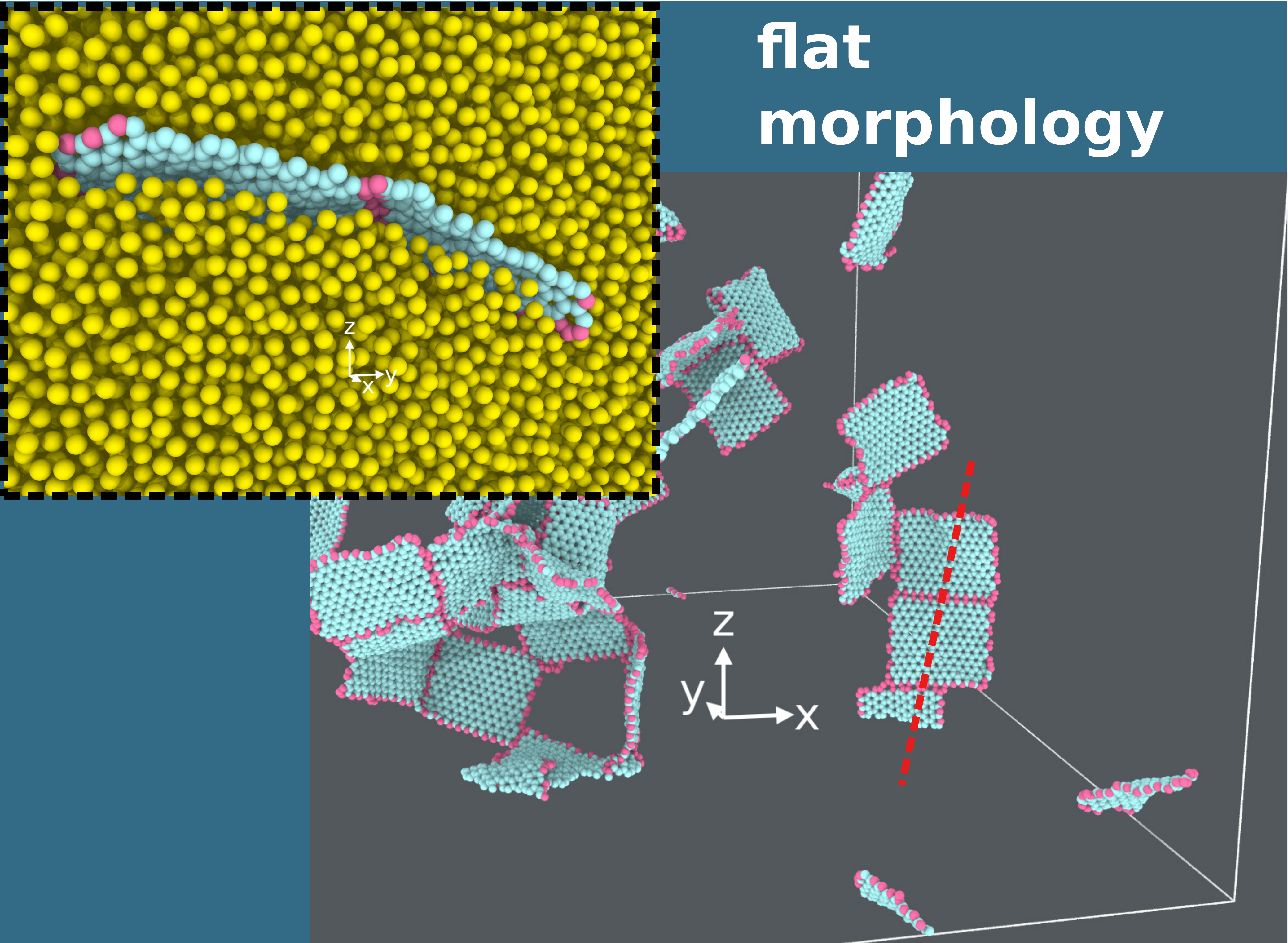}}
\subfigure[]{\includegraphics[width=0.9 \columnwidth]{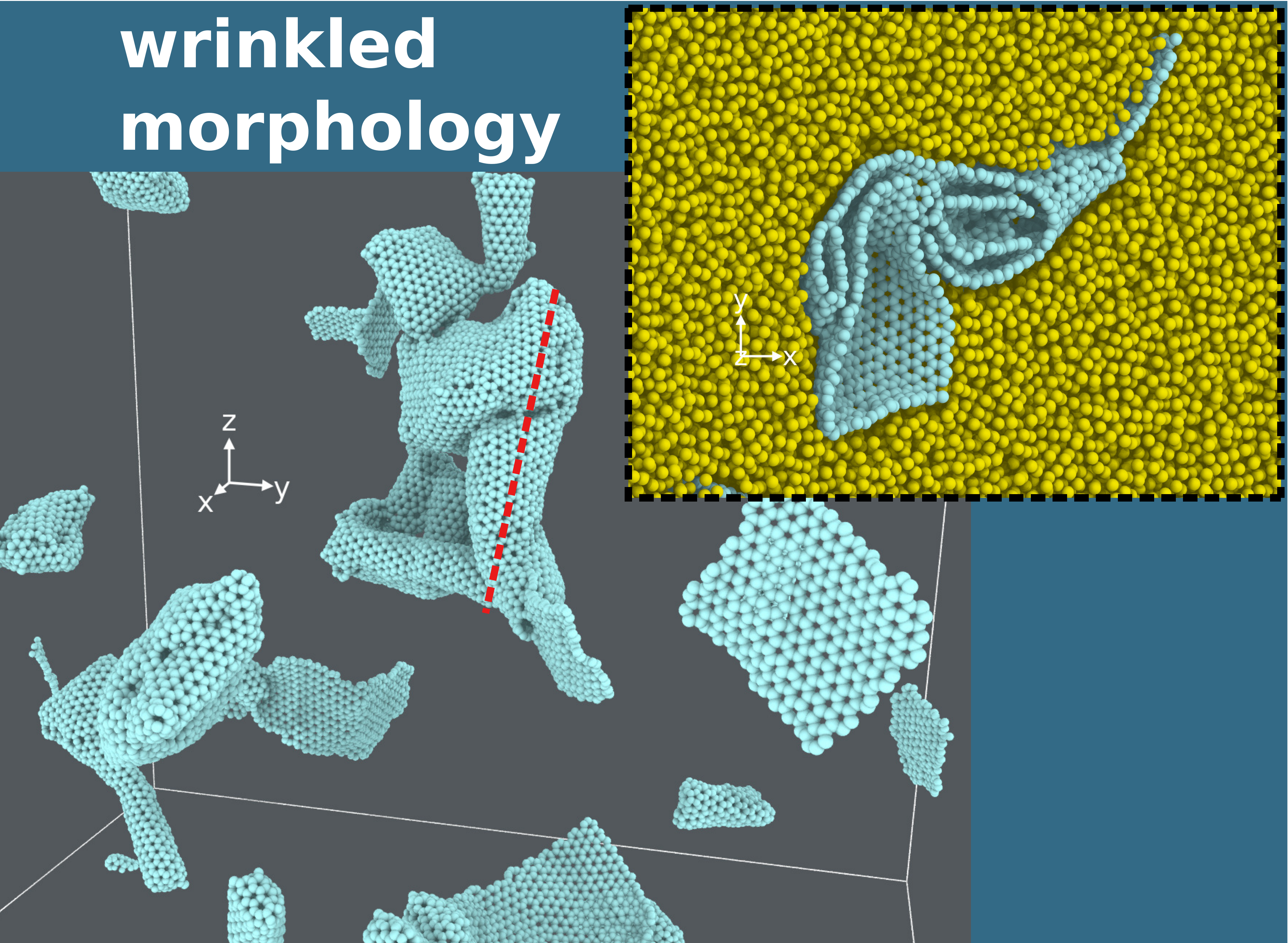}}
\end{center}
\caption{Graphene structures obtained  after equilibration at 2300 K  for (a) graphene flakes with strongly bound edge atoms  and (b) with unmodified edges. 
The red dashed lines show a cut through  the substrate as displayed  in the insets.  Color code as in  \qfig{f_setup}.
}
\label{f_relaxed}
\end{figure}

We let the system equilibrate in an NPT ensemble  at 2300 K for 14 ns. This time is sufficient to equilibrate the systems, see Supplementary Material (SM) \cite{NiGra_SM}.
During the equilibration process, the graphene flakes changed their positions and even their shapes in the liquid metal.
Because of the strong attraction of C atoms modeled by the AIREBO potential \cite{STH00}, graphene flakes tend to aggregate and form covalent bonds between each other. 
However, distinct differences show up between graphene with strongly bound edge atoms  and unmodified edges, see \qfig{f_relaxed}.

\begin{figure}[ht]
\begin{center}

\includegraphics[width=0.9\columnwidth]{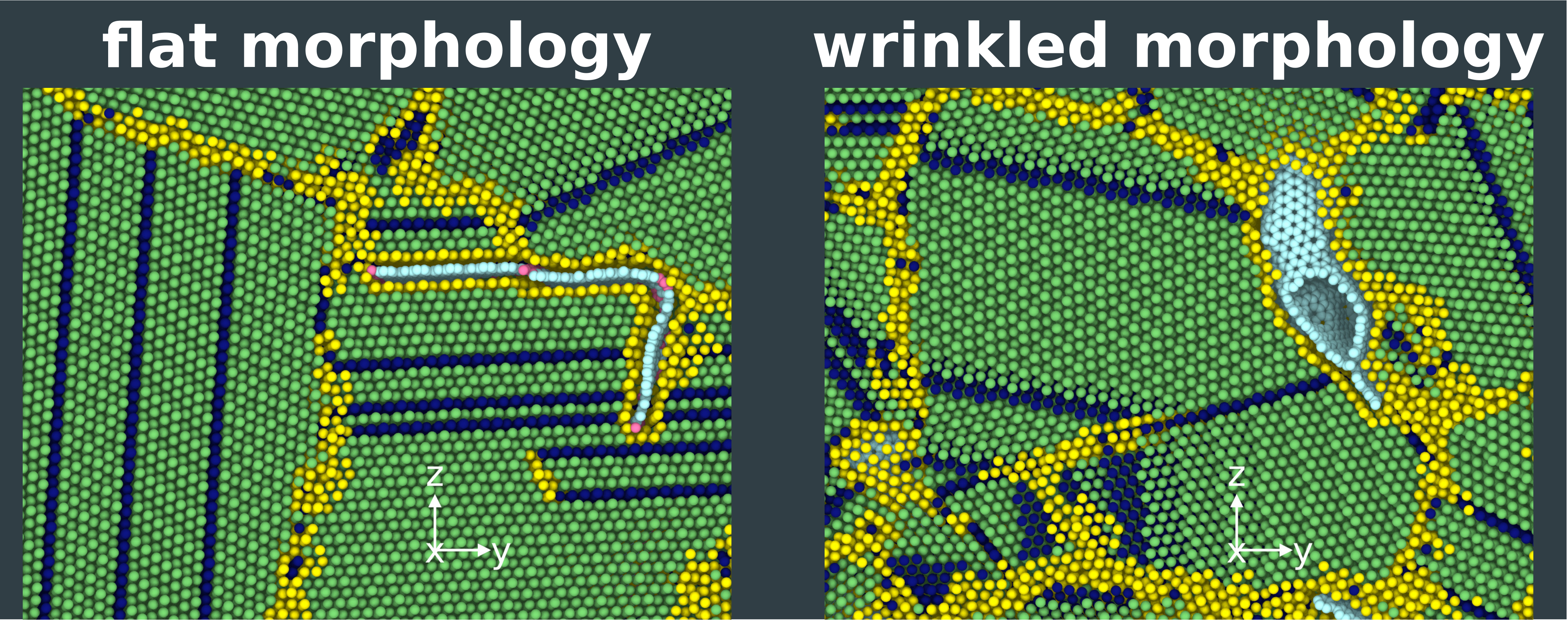}
\end{center}
\caption{Crystallographic structure of Ni-graphene composites after quenching for   (left) graphene flakes with strongly bound edge atoms  and (right) with unmodified edges. 
Atoms are colored according to common-neighbor analysis. Green: fcc nickel; 
dark blue: SFs in nickel; yellow: grain boundaries; light blue: graphene; red: edge atoms of graphene flakes.
The yellow arrow shows the propagation of partial dislocations  leaving behind a twin structure.
}
\label{f_composites}
\end{figure}
In the case of graphene  with strongly bound edge atoms, \qfig{f_relaxed}(a), the  flakes are flat and contact each other at their edges, since the AIREBO potential describing the C-C interaction shows  an increased attraction between C atoms there. 
The final system shows a flat morphology, since the increased C-Ni attraction at the edges prevents further flakes from approaching. 

As we show in the SM, \cite{NiGra_SM}, graphene flakes  with unmodified edges tend to roll up in the liquid Ni metal; the mutual attraction of edge atoms stabilizes the tubular structures. When other flakes approach such a tube, it may attach to the edges resulting in the wrinkled structures we observe. If the Ni-C attraction is increased in strongly bound edge atoms, the flake edges are surrounded by a quite stable Ni shell; this Ni shell reduces the amount of Ni flakes rolling up while not completely erasing it. In addition, it may be supposed that the Ni shell enhances the flexural stiffness of the flakes and thus impedes the roll-up process. As a consequence, we observe a tendency towards flat structures in flakes with  strongly bound edge atoms.

Also in graphene with unmodified edges, the C-C interaction induces the flakes to approach each other at the edges. However, further structural changes occur that are  driven by reactive interactions of the edges of graphene flakes.  The flakes continue further aggregation resulting in a wrinkled morphology with rough surfaces and encapsulated voids.

In experiments, it is known that the structure of graphene flakes inside metal-graphene composites is wrinkled rather than flat \cite{YLZ*13,CJ14,HTL*16,ZZ16,CYW*16,GYG*16,LX17,CZL*16,SWS*20}.

We conclude that by modifying the interaction of graphene edge atoms with Ni, it is possible to modify the morphology of the graphene flakes within the liquid Ni metal. In the following we will discern the  \emph{flat morphology} from the \emph{wrinkled morphology} of composites.

\begin{figure*}[ht]
\begin{center}

\includegraphics[width=0.9\linewidth]{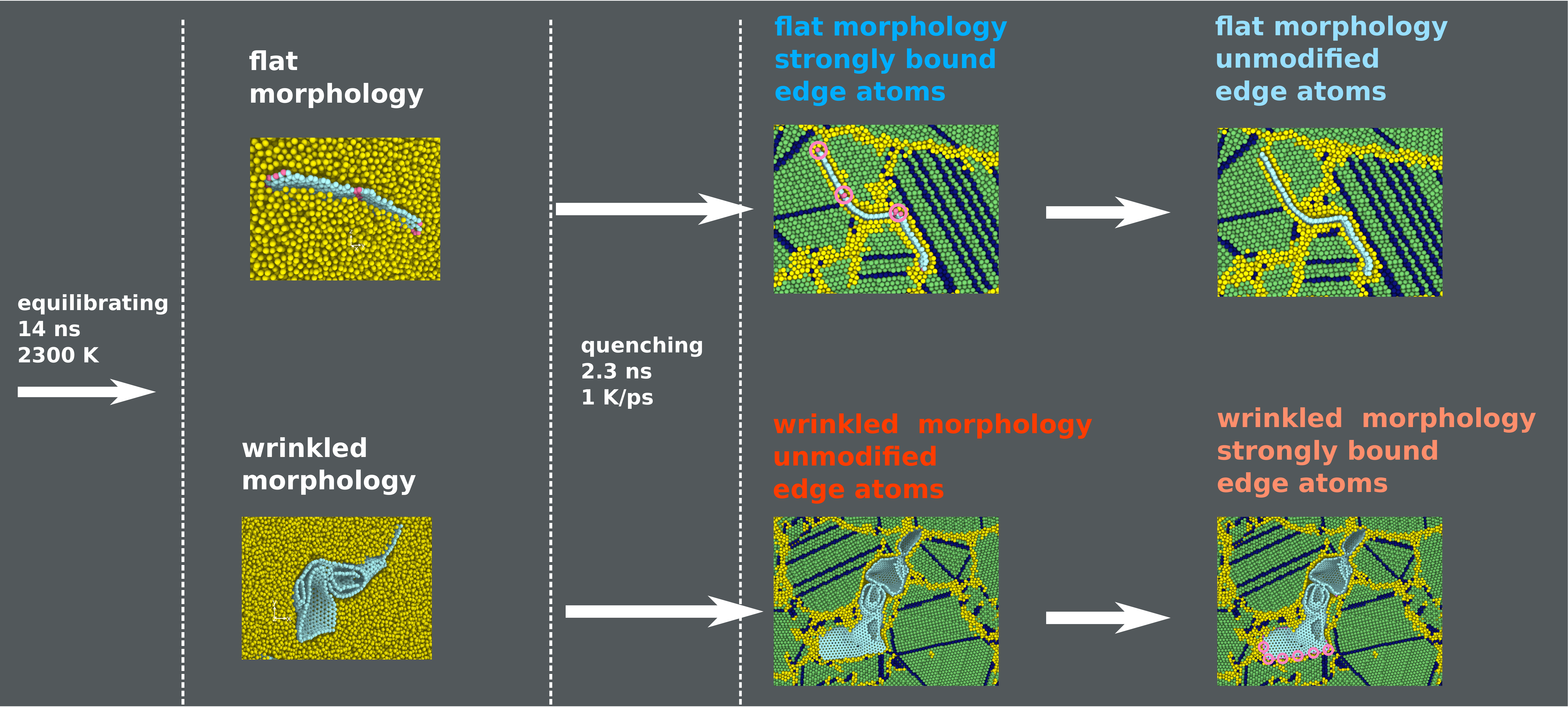}
\end{center}
\caption{Creation of 4 Ni-graphene composites: Two systems with different Ni-graphene interaction  were equilibrated at 2300 K, resulting in a  flat and wrinkled flake morphology, respectively. 
After quenching the composite systems were subjected to  graphene interactions with strongly bound edge atoms and with unmodified edges and then relaxed,  
such that 4 different composite systems were created.
Color code as in  \qfig{f_setup}. 
 }
\label{f_process}
\end{figure*}

\subsection{Quenching}

After equilibration of the liquid, the systems are quenched by decreasing the temperature from 2300 K to 0.1 K with a rate of  1 K/ps. Due to the high quenching rate, Ni does not crystallize as a single crystal, but in a poly-crystalline form, as was discussed in detail for the case of Al in Ref.\   \cite{HDT*16}. We observe an average grain size  of 4.2 (4.8) nm for the composite with wrinkled (flat) morphology.  For pure Ni, which underwent the same quenching procedure, the grain size is comparable, 4.1  nm.
In both composites, the  flake conformation -- wrinkled and flat, respectively -- was conserved during the quenching process, see  \qfig{f_composites}, and the 
graphene flakes are positioned inside the grain boundaries. 
In the case of wrinkled graphene, encapsulated  voids can be found within the graphene, and all quenched systems show a high amount of defects -- in particular stacking fault planes  -- in the grains that originated during the quenching process.

In order to clearly differentiate between the effects of graphene morphology and C-Ni interaction, we create two further systems from these quenched structures by changing the Ni-C interaction of edge atoms and then relaxing the systems.

We thus have the following composite systems:
\begin{enumerate}[(i)]
\item A composite with flat morphology with strongly bound edge atoms.
\item A composite with flat morphology with unmodified edges. It is created from system (i) by changing the interaction of all edge atoms.
\item A composite with wrinkled morphology with unmodified edges.
\item A composite with wrinkled morphology with strongly bound edge atoms.  It is created from system (iii) by  modifying the Ni-C interaction of edge atoms.
\end{enumerate}

These four composites allow us to differentiate the effects of graphene morphology and Ni-C interaction on mechanical properties. 
The generation of these composite systems and their morphologies are summarized in \qfig{f_process}.
\begin{figure}[ht]
\begin{center}
\subfigure[]{\includegraphics[width=0.9\columnwidth]{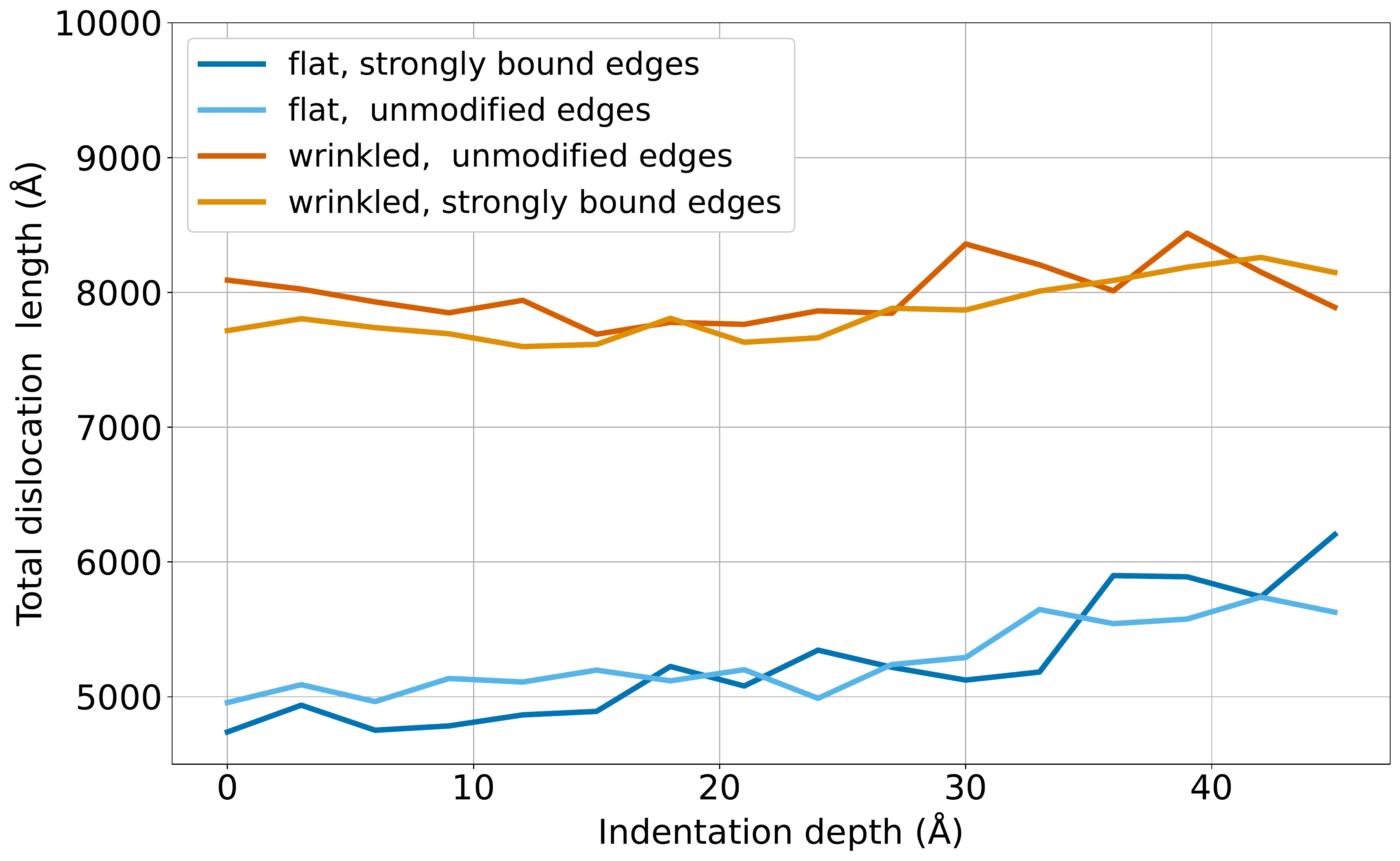}}
\subfigure[]{\includegraphics[width=0.9\columnwidth]{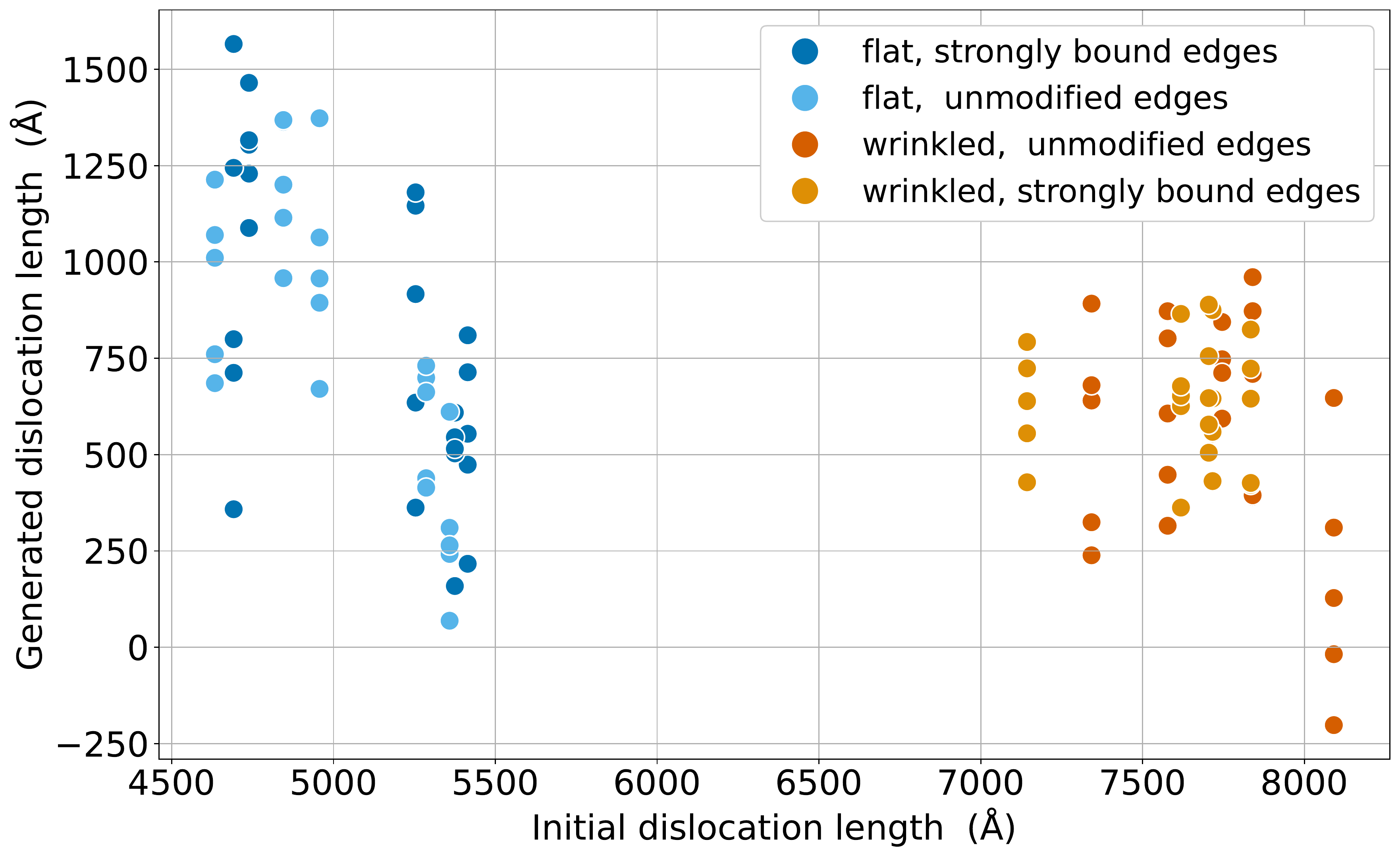}}
\end{center}
\caption{(a) Evolution of the  total dislocation length  with indentation depth. Data are for a representative simulation for each system. 
(b) Dependence of  the length of indentation-induced dislocations  on the initial  dislocation length. Each data point is the result of an individual indentation simulation.
}
\label{f_disloc}
\end{figure}

\subsection{Creation of surfaces}

During equilibration and quenching we apply periodic boundary conditions in all cartesian directions. 
In order to perform indentation simulations, a free surface is created. For increasing statistics, we create 5 replicas of each quenched system and introduce a free surface in each system at a different position.

After introducing free boundary conditions at these surfaces, the nanocomposite systems are  relaxed in the   isothermal-isobaric NPT ensemble.

\subsection{Indentation simulations}

Indentation simulations are performed by modeling the indenter by a repulsive potential  \cite{KPH98},
\begin{equation} \label{e1}
V(r) = \left\{ \begin{array}{ll}
k(R-r)^3, & r<R, \\
0, & r\ge R .
\end{array} \right.
\end{equation}
 The indenter radius is $R=40$ \AA\ and the indenter  stiffness   $k=10$ eV\AA$^{-3}$ \cite{KPH98,ZHU09}.
Using a displacement-controlled algorithm \cite{RBGU17}, 
the indenter is moved perpendicular into the surface to a final depth of 40 \AA\ with a velocity of 20 m/s.

For each system replica, we perform 5 independent indentation simulations which differ from each other by the exact lateral positioning of the indenter; 
it was moved randomly to another position by around $\pm$ 5 \AA. Thus we have 25 indentation simulations for each system. 

The simulations are performed with the open-source code LAMMPS  \cite{Pli95} using a constant time step of 1 fs.
Common-neighbor analysis (CNA) \cite{Stu12} is used to identify the  crystalline  structure and the  dislocation detection algorithm \cite{Stu12,SBA12,SA12} to determine the length of dislocations. 
OVITO \cite{Stu10}  is used to visualize the simulation results.

\section{Results}

\subsection{Dislocation generation}
\begin{figure}[h!]
\begin{center}
\subfigure[]{\includegraphics[width=0.6\columnwidth]{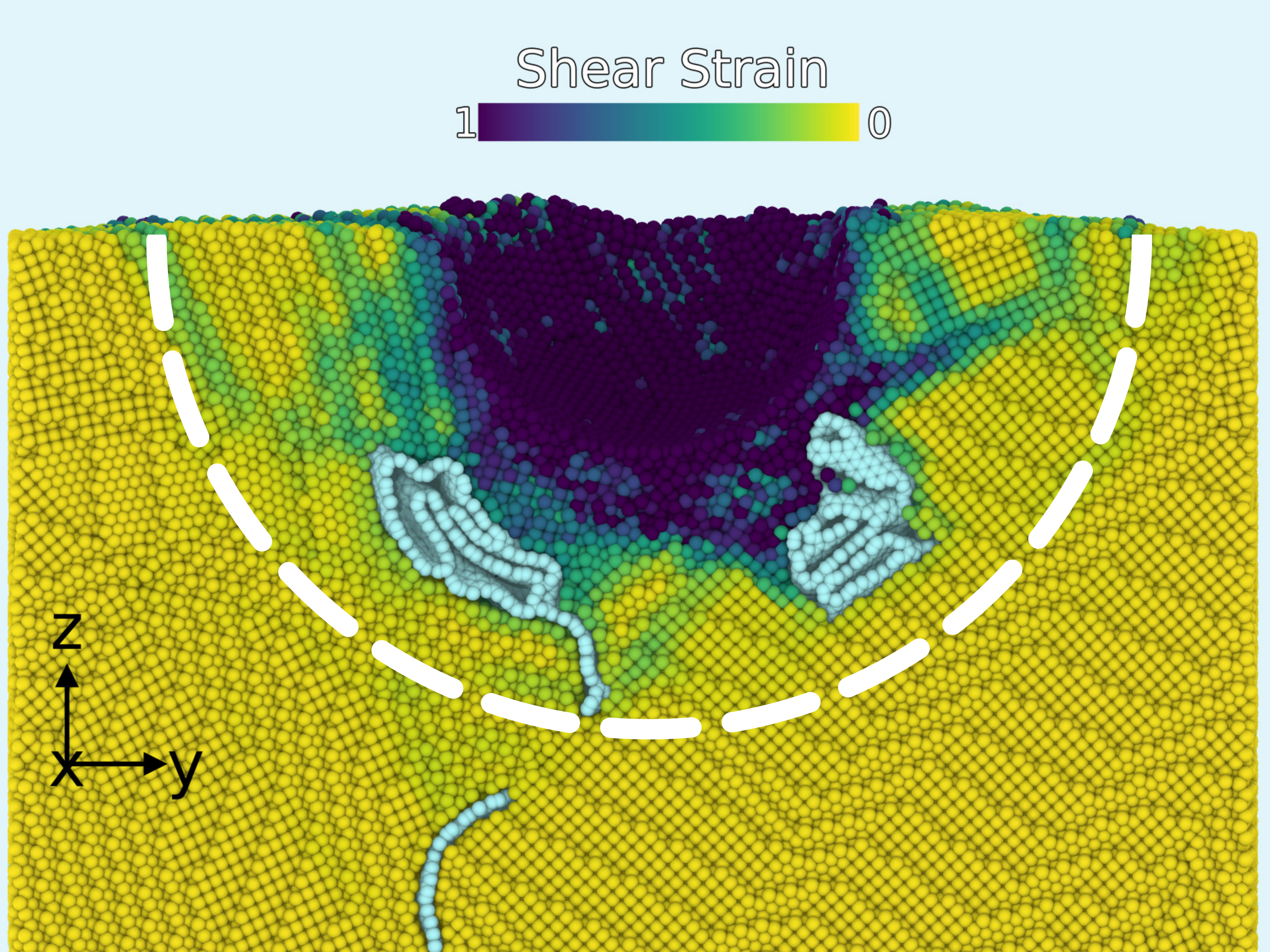}}
\subfigure[]{\includegraphics[width=0.6 \columnwidth]{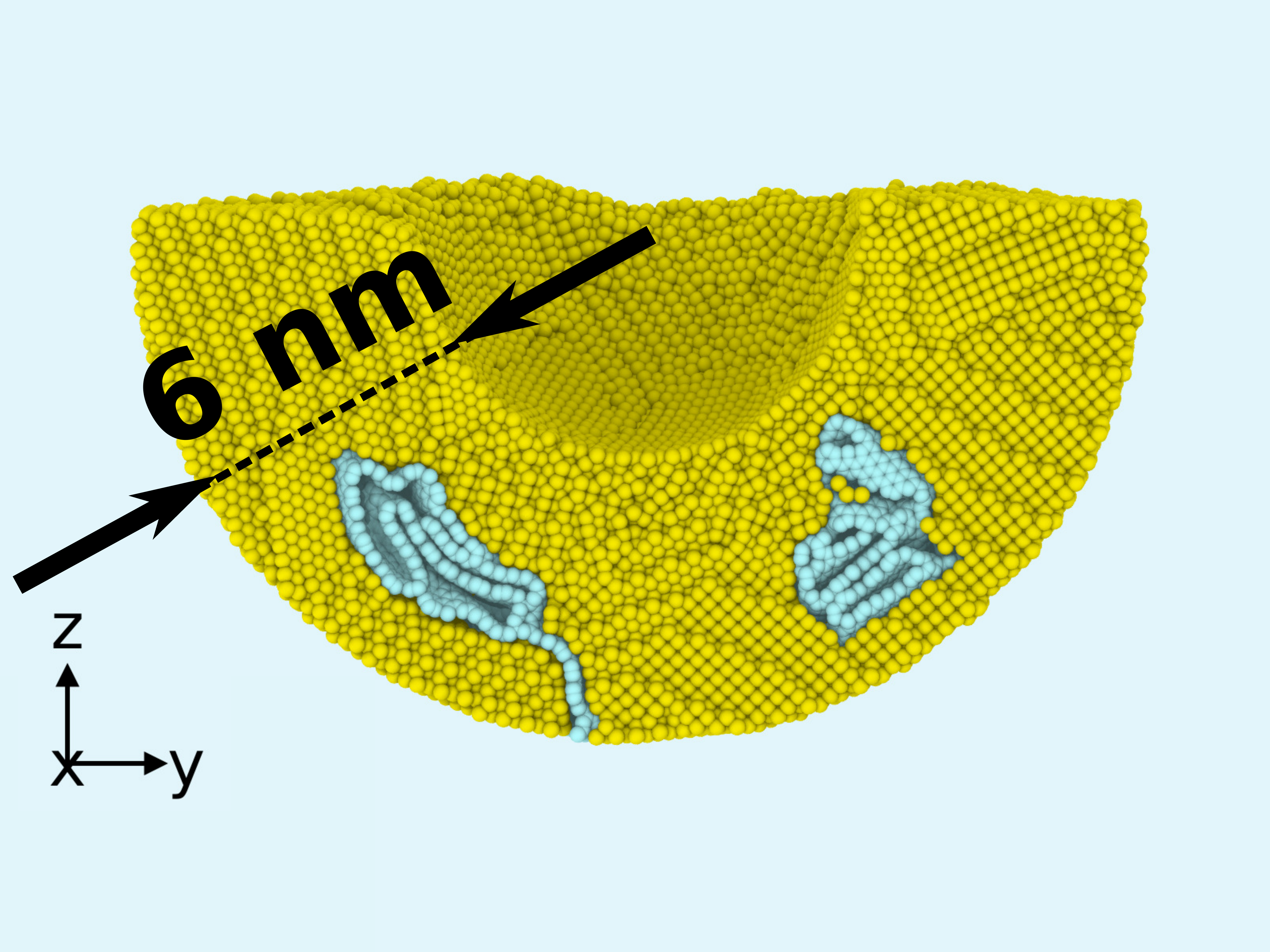}}
\subfigure[]{\includegraphics[width=0.6 \columnwidth]{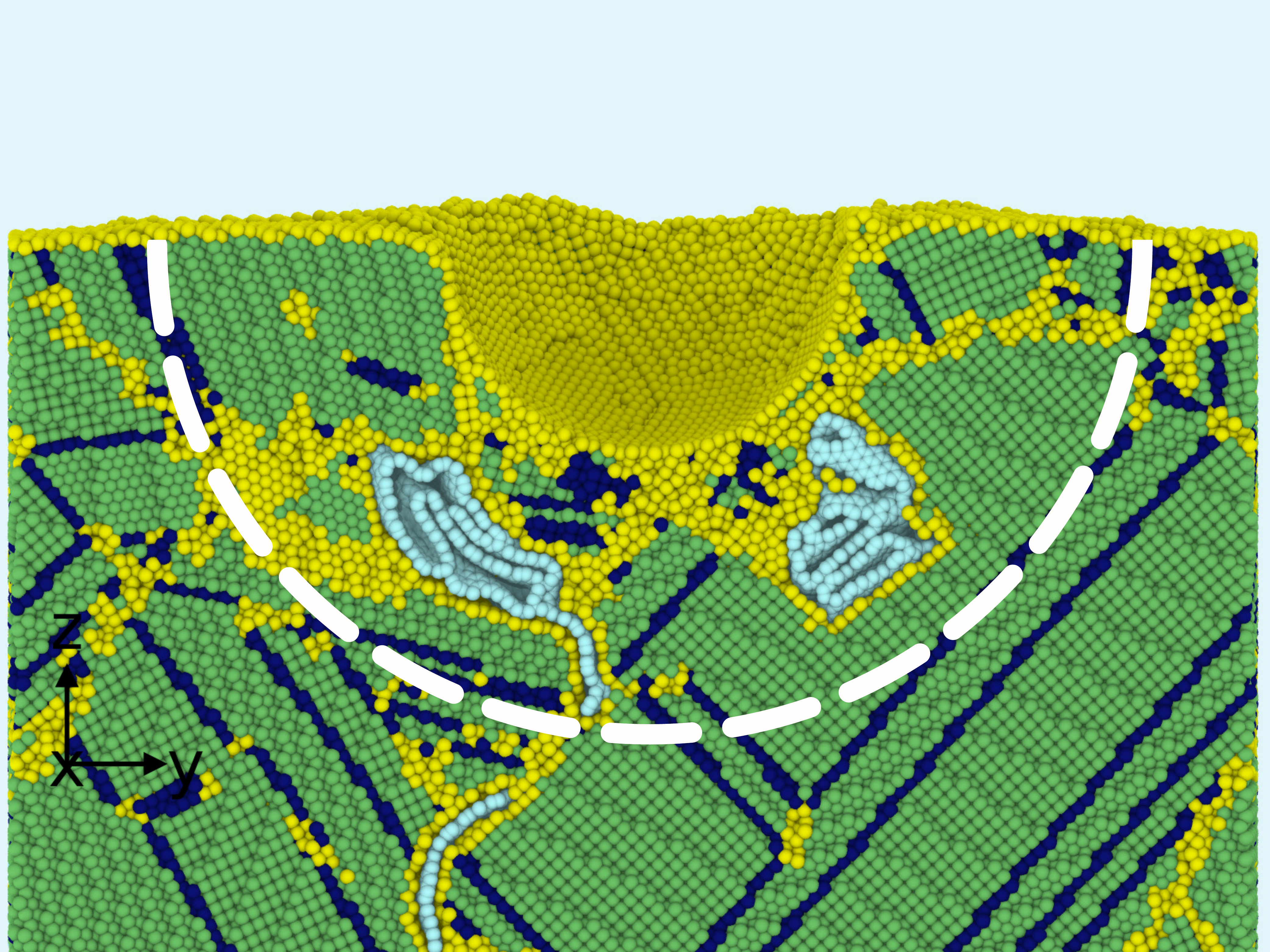}}
\end{center}
\caption{
Side view of  the wrinkled composite with unmodified edges  after indentation.
(a) VMSS map. White dashed line shows the boundary of the plastic zone.
Graphene atoms are colored light blue. Ni atoms are colored according to their VMSS.
(b) Crystallographic structure of this system, where atoms are colored according to common-neighbor analysis. 
 Color code as in  \qfig{f_composites}. 
}
\label{f_shell}
\end{figure}

\begin{figure}[h!]
\begin{center}
\subfigure[]{\includegraphics[width=0.6\columnwidth]{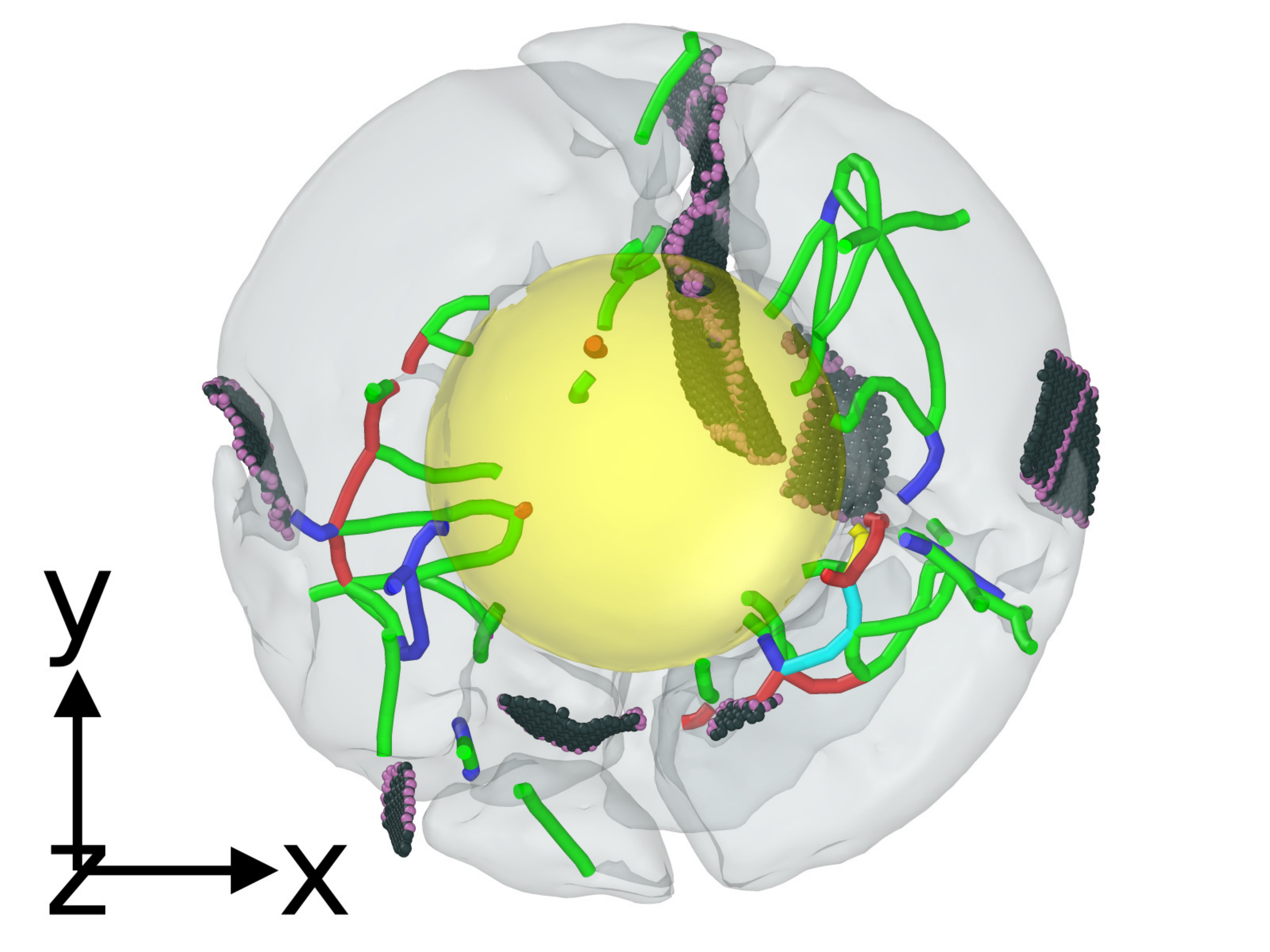}}
\subfigure[]{\includegraphics[width=0.6\columnwidth]{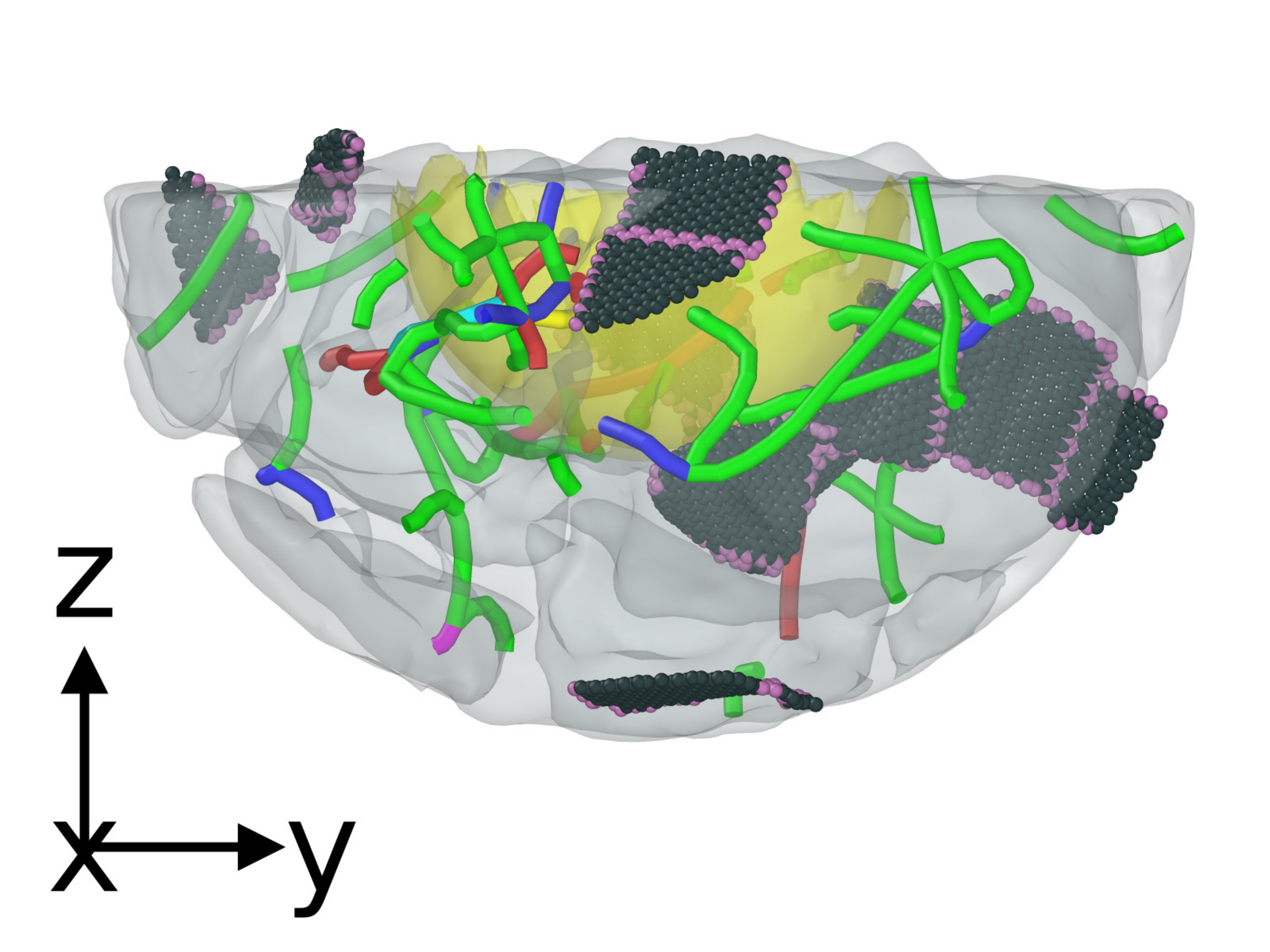}}
\subfigure[]{\includegraphics[width=0.6\columnwidth]{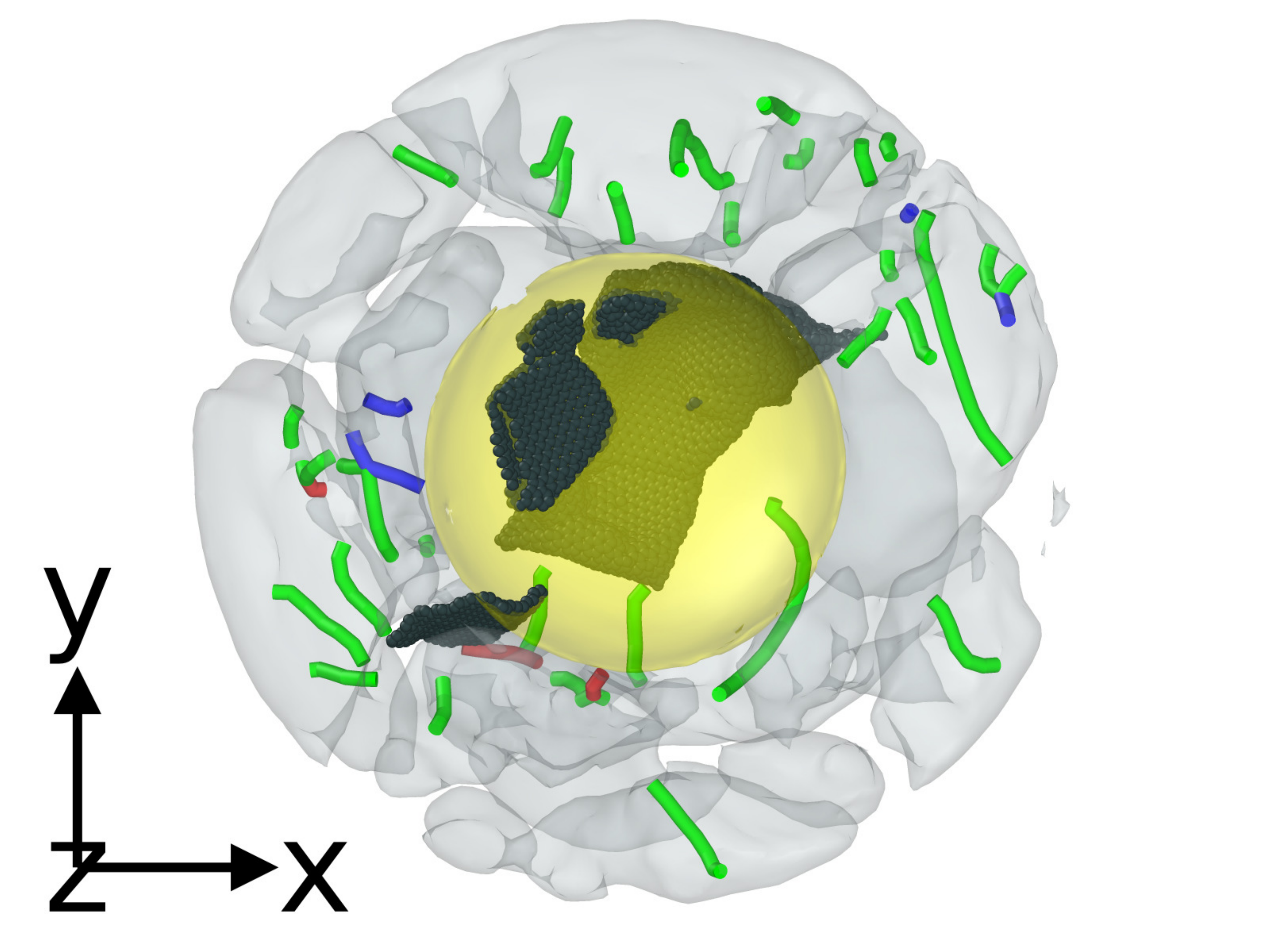}}
\subfigure[]{\includegraphics[width=0.6\columnwidth]{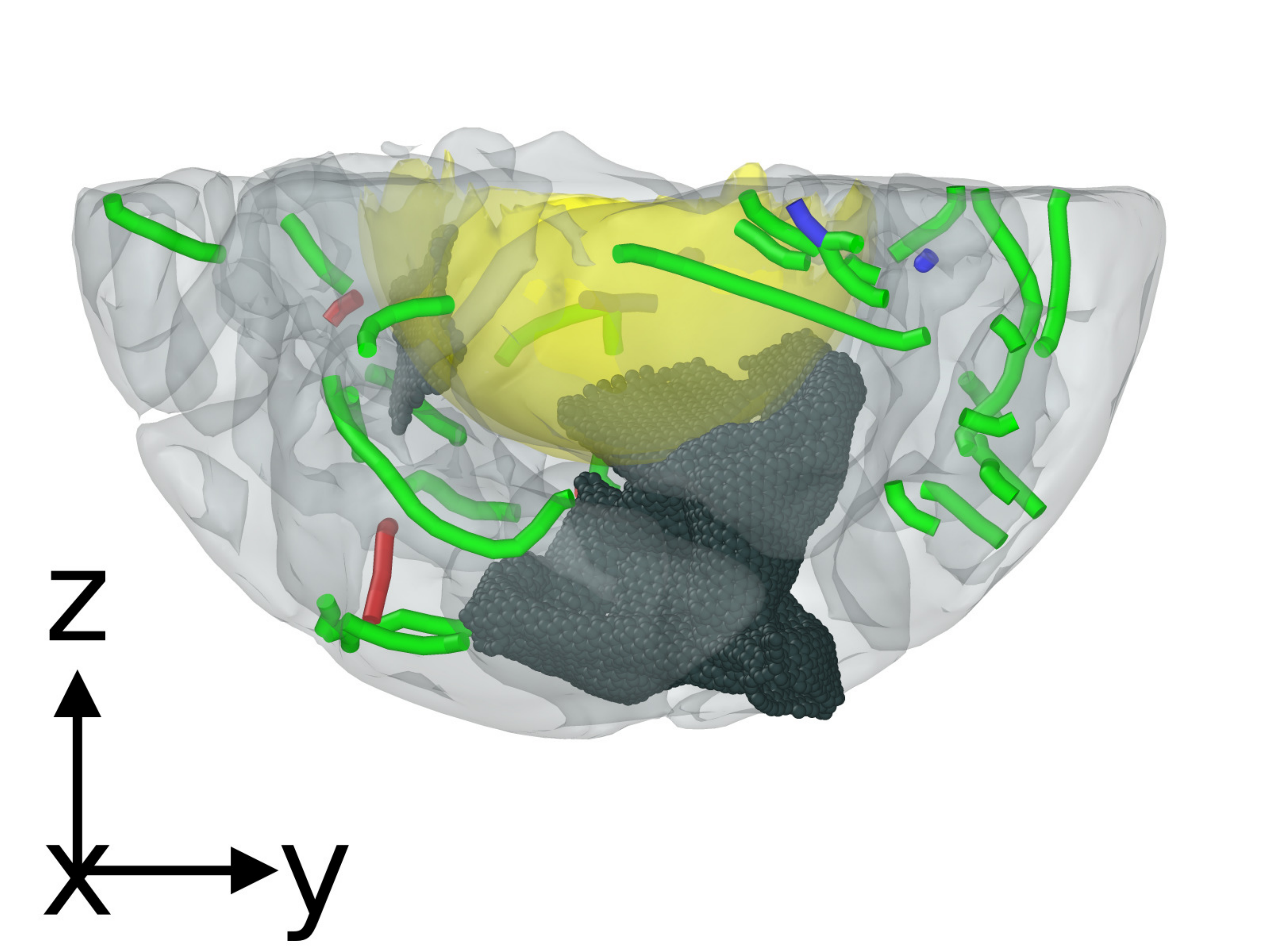}}
\end{center}
\caption{
a) Top view onto the $x$--$y$ plane and  b) side view onto the $x$--$z$ of the plastic zone in the composite with flat morphology and strongly bound edges. 
c) Top view onto the $x$--$y$ plane and d) side view onto the $x$--$z$ of the plastic zone in  the composite with wrinkled morphology and unmodified edges. 
Dislocations are colored according to their Burgers vector. 
Green: $\frac{1}{2}\langle112\rangle$;  dark blue: $\frac{1}{2}\langle110\rangle$; pink: $\frac{1}{6}\langle110\rangle$;  yellow: $\frac{1}{3}\langle001\rangle$; red: other.
Yellow shading shows indented surface. Gray contours highlight Ni grains. 
}
\label{f_shell_disloc}
\end{figure}

\qfig{f_disloc}(a) shows the evolution of the total dislocation  length  during indentation for representative cases. 
In the following, we consider only dislocations generated inside the grains  and leave out those inside the grain boundaries or very close to them. As a result of the solidification of   composites, all systems already 
 have a  dislocation network before indentation, which we denote as `initial dislocations'.  
During indentation, new dislocations are nucleated mainly from the  grain boundaries as well as from high-stress regions in the grains close to the indenter; this can be seen in \qfig{f_disloc}(a) as a general steady  increase of the total dislocation length with increasing indentation depth. This increase is dominant in the case of the flat graphene morphology, while the wrinkled morphology also shows strong reductions in the dislocation length.
Unlike in the case of single-crystalline metals, where the total dislocation length exhibits a  characteristic steady increase \cite{VU19},
here  the absorption of dislocations by grain boundaries and graphene flakes  acts to reduce the  dislocation length.
 \qfig{f_disloc}(a) thus shows 
that  the wrinkled morphology  is more effective in absorbing dislocations. 

 \qfig{f_disloc}(b) shows the dependence of the length of generated dislocations on the initial dislocation length for all 100 simulations. Due to the various surfaces cut out of the quenched composite system and the inaccuracies of the dislocation detection algorithm \cite{Stu12,SBA12,SA12}, the length of the initial dislocations shows some spread. The spread in the generated dislocations is even larger, since for each generated surface, 5 random indentation points were chosen and the generated plasticity depends on the exact  indentation point. 
 Here and in the following, we determine the length of generated dislocations by the difference between the  final and initial length of dislocations for each system.
Note that the length of generated dislocations does not correlate with the initial dislocation length.
On average, systems with a flat flake morphology  show an increase  of the dislocation network by on average 80 nm, while in the case of the   wrinkled morphology  the increase is smaller, 58 nm. In selected cases, a wrinkled geometry may even lead to a reduction of the dislocation network due to the strong absorption activity of the wrinkled flakes. In no cases, an effect of the C-Ni interaction of the edge atoms on the length of the generated dislocations could be observed.

The plastic deformation  is mapped by the von-Mises shear strain (VMSS)  in \qfig{f_shell} (a).
The plastic deformation of a polycrystalline sample is a   consequence of dislocation nucleation inside the grains and 
shear activation in grain boundaries. We identify the boundary of the plastic zone in this figure
such that all shear strains are situated in the plastic zone. We find that in all composites studied, the radius of the plastic zone amounts to roughly 10--15 nm.
\qfig{f_shell} (b) displays a CNA analysis of the same system as in \qfig{f_shell} (a). It shows that 
the two wrinkled graphene flakes existing in the plastic zone create large defective regions inside the  plastic zone.

The dislocation  structure in the grains after indentation is shown in \qfig{f_shell_disloc}  for two different composites.


\begin{itemize}
\item \qfig{f_shell_disloc} (a,b) shows the dislocation structure in the case of the flat morphology with strongly bound edge  atoms.
In this particular simulation, there are 2690 C atoms inside the plastic zone. The total length of dislocations amounts to  1290 \AA.
This large value indicates pile-up of dislocations inside the plastic zone.
Dislocations can be seen around the graphene flakes. Thus, the  graphene-nickel interface in case of flat morphology is an additional barrier for dislocations.  

\item \qfig{f_shell_disloc} (c,d) shows the dislocation structure in the case of the  wrinkled morphology with unmodified edges.
There are 6301 C atoms inside the plastic zone and the total length of dislocations amounts to 540 \AA.
This morphology results in a more inhomogeneous dispersion of graphene inside the plastic zone.
Because graphene is not flat, the interface between graphene and nickel has a higher tendency of absorbing dislocations.
Dislocation nucleation occurs only  in grains far from graphene.

\end{itemize}

\begin{figure*}[ht]
\begin{center}
\subfigure[]{\includegraphics[width=0.8\columnwidth]{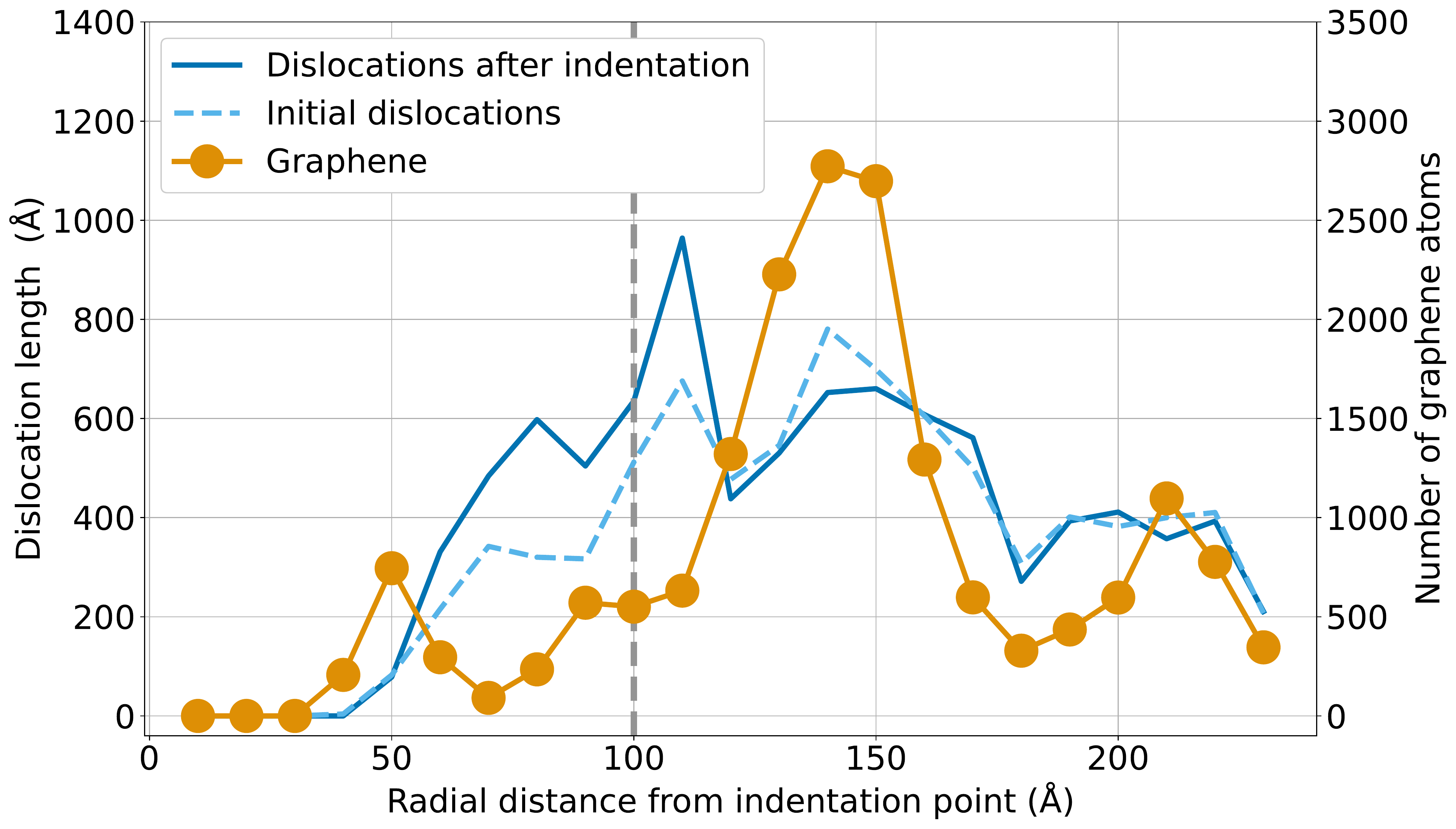}}
\subfigure[]{\includegraphics[width=0.8\columnwidth]{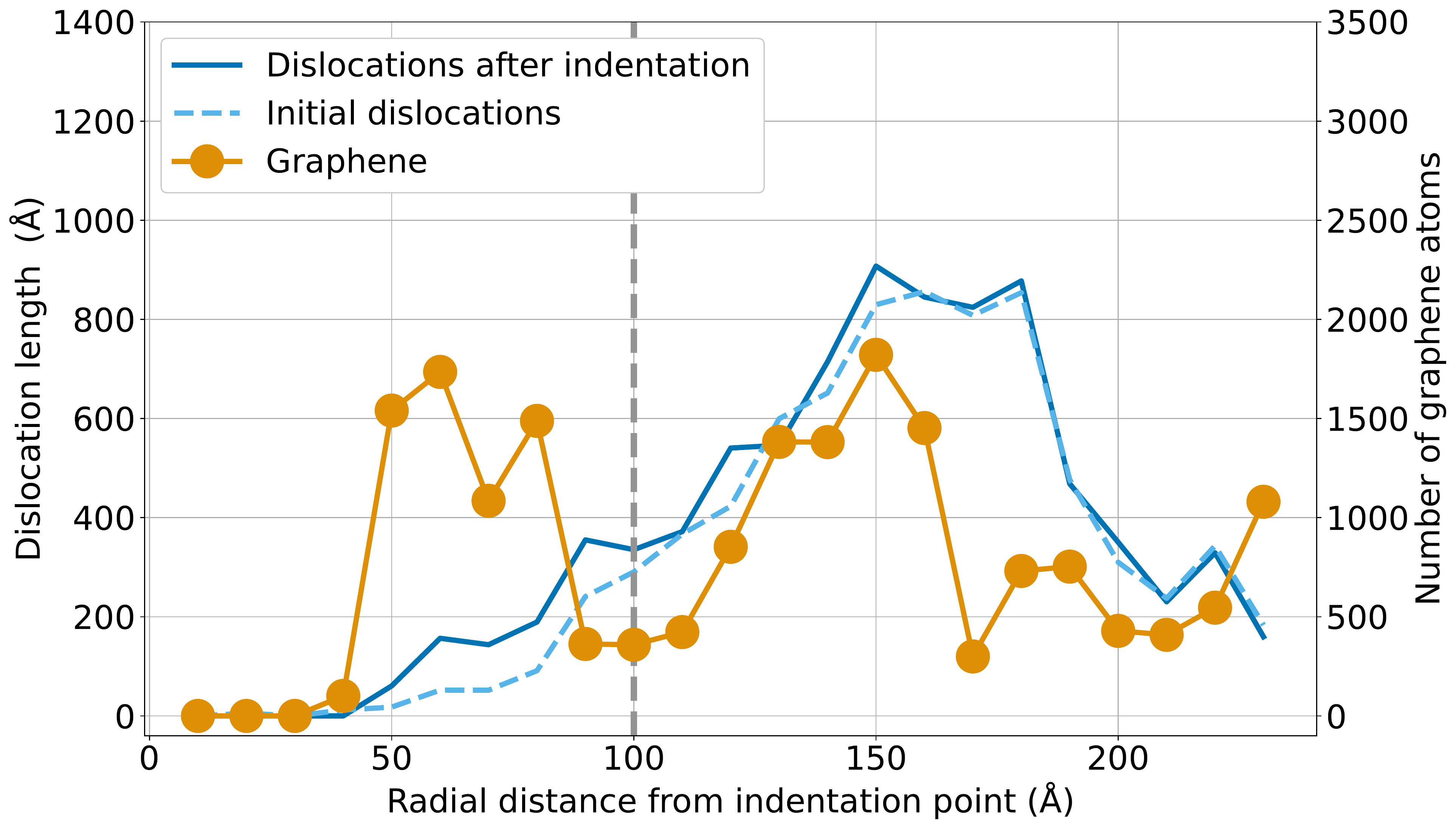}}

\end{center}
\caption{
Radial dependence of the initial and final dislocation length as well as  the number of carbon atoms 
for (a) the flat morphology with strongly bound edges and (b) the wrinkled morphology with unmodified edges. 
Data show quantities in radial shells of 1 nm width around the indentation point.
The vertical dashed line indicates the extension of the plastic zone.  
}
\label{f_correlation}
\end{figure*}

\qfig{f_correlation} provides quantitative details about the dependence of the dislocation generation and the distribution of graphene in the plastic zone and around it. To this end we determine the length of dislocations before and after indentation in spherical shells of width 1 nm around the indentation point. Two scenarios have been selected.
 \qfig{f_correlation}a shows a case, where the amount of graphene within the plastic zone is small, but more massive outside, at a distance of 15 nm from the indentation point. As a consequence, a considerable increase of the dislocation length in the plastic zone and a pile-up of dislocations in front of the graphene peak at 15 nm can be observed. In the second case, \qfig{f_correlation}b, the amount of graphene within the plastic zone is high, and not so massive outside of the zone. Here the length of dislocations within the plastic zone is reduced during indentation; this plot directly shows the effectivity of graphene to absorb dislocations.



We conclude that the dislocations that are invariably created during the indentation are absorbed by the graphene flakes; a large amount of graphene in the plastic zone may even reduce the initial length of dislocations. The effect of the flake morphology shows up by a higher absorption activity of wrinkled flakes compared to flat flakes. The effect of C-Ni interaction on dislocation activity is negligible.

\subsection{Hardness}

\begin{figure*}[ht]
\begin{center}
\subfigure[]{\includegraphics[width=0.95\columnwidth]{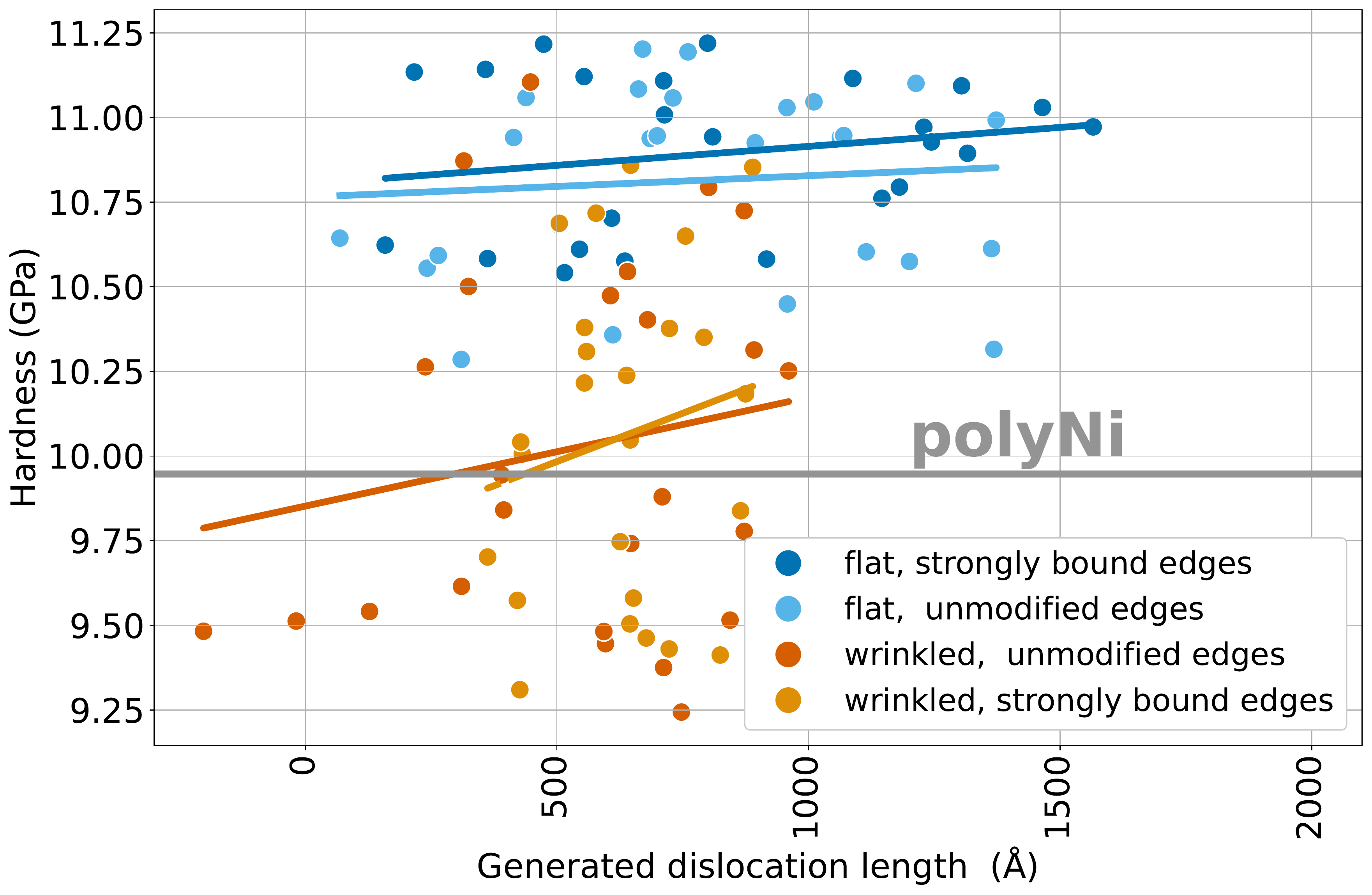}}
\subfigure[]{\includegraphics[width=0.95\columnwidth]{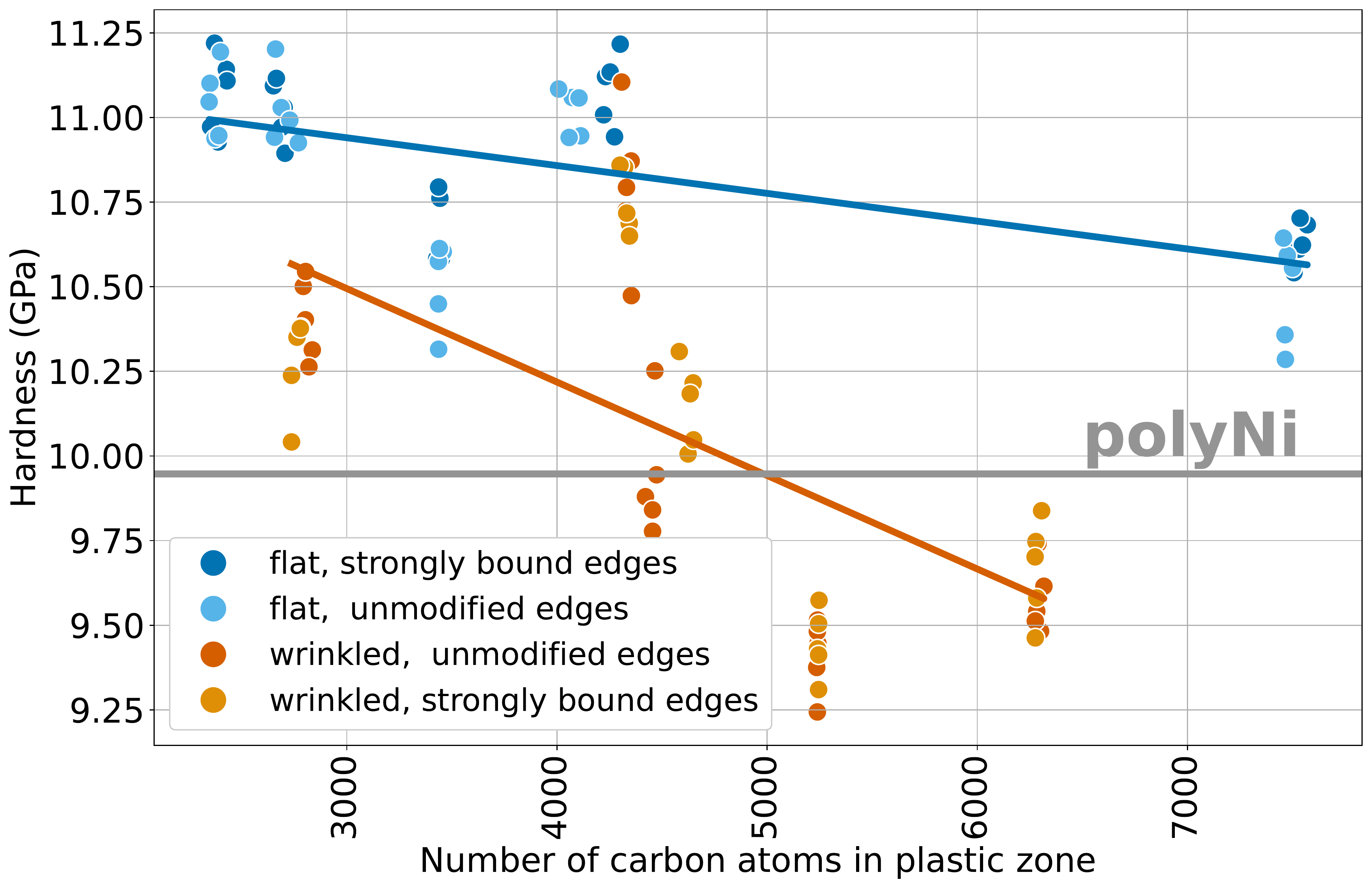}}
\subfigure[]{\includegraphics[width=0.95\columnwidth]{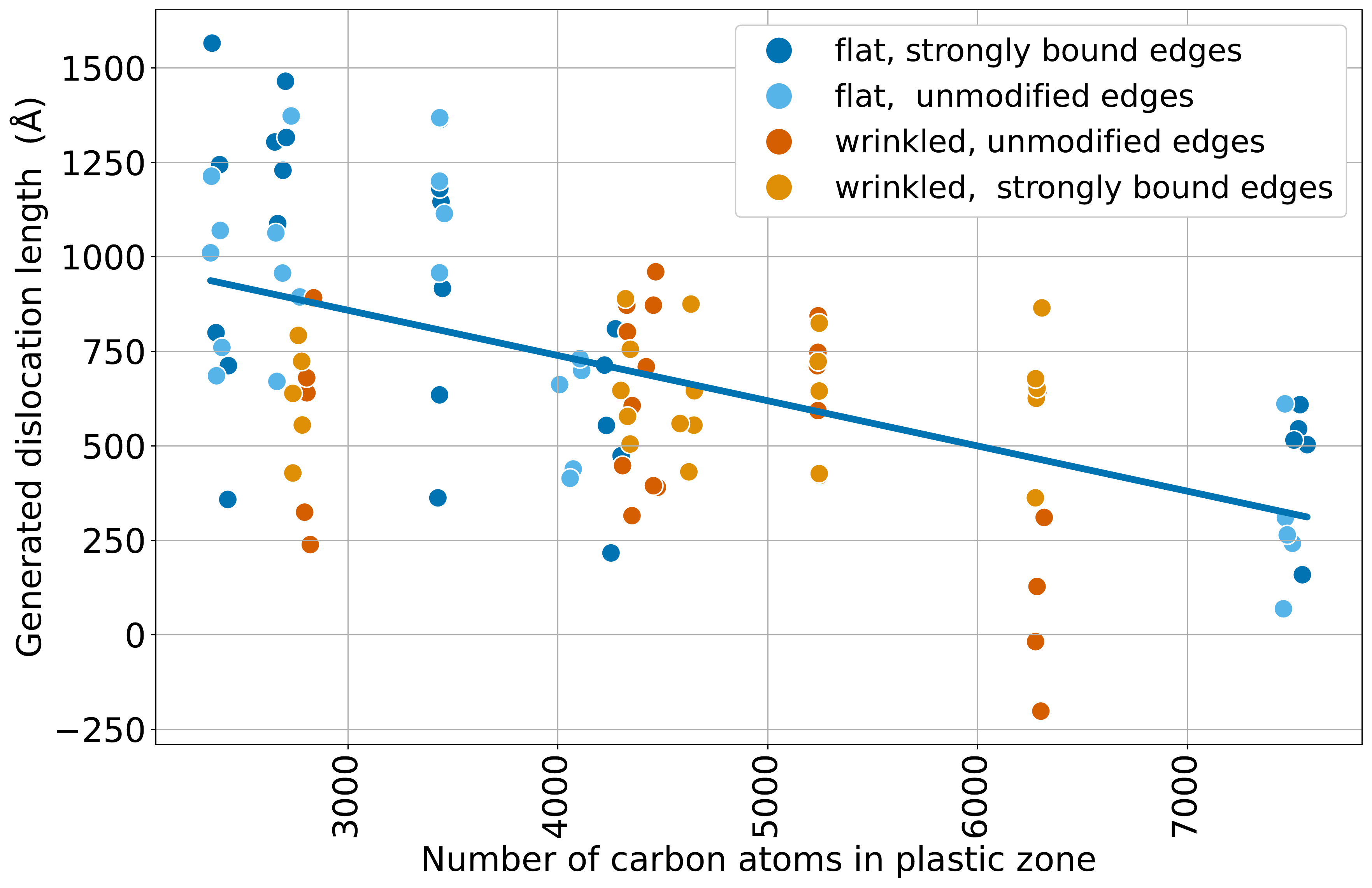}}
\end{center}
\caption{(a) Dependence of composite hardness on  (a) the length of  dislocations generated in grains and  
(b) the  number of carbon atoms in the plastic zone.  
(c) Dependence of the length of  dislocations generated in grains on the number of carbon atoms in the plastic zone. Lines are to guide the eye. The gray line gives the average hardness of polycrystalline Ni. 
}
\label{f_hardness}
\end{figure*}

Hardness was determined as the contact pressure averaged over indentation depths between 20 \AA\ and
40 \AA, where the hardness already showed its stable, saturated value, see SM \cite{NiGra_SM}. The results of all 100 indentations are  shown in \qfig{f_hardness}. 
\qfig{f_hardness} (a) shows the dependence of the hardness on the length of generated dislocations. Even though the data show a  considerable scatter, several tends can be observed. Thus, a common trend is seen in that all systems show a steady increase of hardness upon increasing of dislocation length. This is caused by dislocation pileup which increases hardness \cite{MHH05,WM11,BDMU15}. 

Polycrystalline nickel -- with a grain size of 4.1 nm -- shows a hardness of 9.95 GPa. This is in reasonable agreement with literature data \cite{LYW13} that  obtained around 12 GPa for 5-nm grained polycrystalline Ni; the slight difference may be explained by the different interatomic interaction potentials used.

Composites with flat morphology show an enhanced hardness as compared to the wrinkled morphology, see \qfig{f_hardness}a; the difference amounts to around 1 GPa on average. The hardening of polycrystalline Ni by flat flakes is particularly well seen by comparing the hardness results with those of  polycrystalline Ni:  the hardness of teh flat morphology is in all cases above that of polycrystalline nickel. This means a definite strengthening of the composite. The reason lies in the way how the 2D graphene flake can sustain the stress field exerted by the indenter on the composite material.  In the case of the flat morphology, in-plane deformation of the flake will result in  hardening, while in the case of the wrinkled morphology, in addition out-of-plane deformations will assist to sustain  the load. As the stiffness of 2D materials is considerably stronger for in-plane stresses than for out-of-plane stresses, the flat material will be more effective in sustaining these stresses and thus enable a hardening of the composite as seen in our \qfig{f_hardness}a.


In general, wrinkled graphene shows less resistance against deformation than flat graphene. Nicholl \etal 
 \cite{NCL*15} find that the in-plane stiffness of graphene is strongly reduced upon crumpling.
In our simulations, we find that wrinkled graphene also more easily deforms during loading of the composite  resulting in out-of-plane deformations and thus offering an additional mode of reducing the hardness of the composite (see SM, \cite{NiGra_SM}, for an example). In addition, wrinkled graphene allows for a similar or even better way to dislocation absorption than flat graphene.


The effect of flat graphene on the mechanical properties of Ni has been repeatedly studied in the past \cite{CNB13,MSN18,LWWW16,YBM17,VU19,VU20,SA21,SDA21,ZHW20}. There it has been shown that graphene mainly absorbs dislocations and hence weakens the nickel matrix as compared to a single-crystalline Ni block whose hardness amounts to 15 GPa \cite{VU19}. 
Yazdandoost \etal \cite{YBM17} investigated the effect of graphene flakes inserted into the grain boundaries of a polycrystalline Ni matrix and found almost no effect on hardness, arguing that dislocation nucleation and blocking cancel in the case of a random orientation of the flat graphene flakes.  In the present work, we find that flat graphene flakes embedded in the Ni grain boundaries harden the polycrystalline Ni matrix. We presume that this occurs because our Ni grain boundaries and the enclosed flakes have been carefully relaxed in our sample preparation, \qsect{s_m},
in contrast to the procedure based on Voronoi tesselation in Ref.\ \cite{YBM17}. 
In the case of wrinkled flakes, besides the obvious effect of dislocation absorption which weakens the matrix, the out-of-plane deformation of the flakes adds a second effect such that the overall hardness of the wrinkled composite lies below that of polycrystalline Ni. 
The present work shows that the morphology of graphene affects the hardness of the composite in a more refined manner, such that both weakening and softening of the matrix material becomes possible. 




\qfig{f_hardness} (b) shows the dependence of the composite  hardness on the number of carbon atoms in the plastic zone.
In all cases, the  hardness decreases with increasing graphene content. Thus, graphene flakes are not per se strengthening the composite. Rather, graphene flakes act as defect structures that   absorb dislocations.

\qfig{f_hardness} (c) shows the  effect of graphene atoms on dislocation formation inside the plastic zone. An increase of the  graphene concentration leads on average to less dislocation generation in the plastic zone; this trend is visible for all systems studied. It is caused by the absorption of dislocations by graphene flakes in the plastic zone.

In all our correlations, \qfigs{f_hardness}a--c, no difference in hardness based on the Ni-C edge atom interaction shows up. In all cases, flakes with strongly bound edges and unmodified edge atoms feature on average the same hardness. 
We thus conclude that while the effect of the graphene morphology on the mechanical properties is strong, the effect of the C-Ni interaction of the edge atoms itself is minor.


\section{Summary}

Using molecular dynamics simulation, we studied the effect of the morphology of graphene flakes in a Ni-graphene nanocomposite on the mechanical properties, in particular the hardness. The flake morphology was established by the self-organization of graphene flakes in a liquid Ni matrix. By varying the  interaction potential of graphene edge atoms with Ni, two different flake morphologies -- flat vs wrinkled -- could be generated. The modification of the interaction potential is justified by the fact that edge atoms have unsaturated bonds, in contrast to interior flake atoms, and are therefore prone to generating strong Ni-C bonds. In the quenched solid structures of the Ni-graphene nanocomposites, graphene is located in the grain boundaries of Ni.

The mechanical properties of the composite were evaluated in a simulated nanoindentation test. As known from previous studies \cite{MHH05,WM11,BDMU15}, the dislocations generated by the indentation process will pile up in front of the Ni grain boundaries and the graphene flakes, leading to a hardening of the composite compared to a pure poly-crystalline Ni specimen. We could verify by selected case studies that a high concentration of graphene in the plastic zone of the nanoindenter leads to a strong absorption of dislocations, such that even some of the dislocations present by the solidification process before the indentation were annihilated.  
However, this effect only occurs for the wrinkled flake morphology; there, the absorption of dislocations at the flake dominates the pile-up and leads to a strong reduction in hardness, amounting to roughly10 \% in the case investigated. 

While the effect of the  flake morphology on the hardness is strong,  the changed interaction between graphene edge atoms and the Ni matrix shows only negligible effects on the dislocation activity and hardness. Our results thus suggest that future studies of mechanical properties of metal-graphene composites need to take the flake morphology in due account.



\begin{acknowledgments}
We acknowledge support by the Deutsche Forschungsgemeinschaft (DFG, German Research Foundation) -- project number 252408385 --  IRTG 2057.
Simulations were performed at the High Performance Cluster Elwetritsch (RHRK, TU Kaiserslautern, Germany). 
\end{acknowledgments}

\bibliography{./bibo/STRING.bib,./bibo/all.bib,./bibo/PUBL.bib}

\begin{thebibliography}{82}%
\makeatletter
\providecommand \@ifxundefined [1]{%
 \@ifx{#1\undefined}
}%
\providecommand \@ifnum [1]{%
 \ifnum #1\expandafter \@firstoftwo
 \else \expandafter \@secondoftwo
 \fi
}%
\providecommand \@ifx [1]{%
 \ifx #1\expandafter \@firstoftwo
 \else \expandafter \@secondoftwo
 \fi
}%
\providecommand \natexlab [1]{#1}%
\providecommand \enquote  [1]{``#1''}%
\providecommand \bibnamefont  [1]{#1}%
\providecommand \bibfnamefont [1]{#1}%
\providecommand \citenamefont [1]{#1}%
\providecommand \href@noop [0]{\@secondoftwo}%
\providecommand \href [0]{\begingroup \@sanitize@url \@href}%
\providecommand \@href[1]{\@@startlink{#1}\@@href}%
\providecommand \@@href[1]{\endgroup#1\@@endlink}%
\providecommand \@sanitize@url [0]{\catcode `\\12\catcode `\$12\catcode
  `\&12\catcode `\#12\catcode `\^12\catcode `\_12\catcode `\%12\relax}%
\providecommand \@@startlink[1]{}%
\providecommand \@@endlink[0]{}%
\providecommand \url  [0]{\begingroup\@sanitize@url \@url }%
\providecommand \@url [1]{\endgroup\@href {#1}{\urlprefix }}%
\providecommand \urlprefix  [0]{URL }%
\providecommand \Eprint [0]{\href }%
\providecommand \doibase [0]{http://dx.doi.org/}%
\providecommand \selectlanguage [0]{\@gobble}%
\providecommand \bibinfo  [0]{\@secondoftwo}%
\providecommand \bibfield  [0]{\@secondoftwo}%
\providecommand \translation [1]{[#1]}%
\providecommand \BibitemOpen [0]{}%
\providecommand \bibitemStop [0]{}%
\providecommand \bibitemNoStop [0]{.\EOS\space}%
\providecommand \EOS [0]{\spacefactor3000\relax}%
\providecommand \BibitemShut  [1]{\csname bibitem#1\endcsname}%
\let\auto@bib@innerbib\@empty
\bibitem [{\citenamefont {Xiong}\ \emph {et~al.}(2015)\citenamefont {Xiong},
  \citenamefont {Cao}, \citenamefont {Guo}, \citenamefont {Tan}, \citenamefont
  {Fan}, \citenamefont {Li},\ and\ \citenamefont {Zhang}}]{XCG*15}%
  \BibitemOpen
  \bibfield  {author} {\bibinfo {author} {\bibfnamefont {D.-B.}\ \bibnamefont
  {Xiong}}, \bibinfo {author} {\bibfnamefont {M.}~\bibnamefont {Cao}}, \bibinfo
  {author} {\bibfnamefont {Q.}~\bibnamefont {Guo}}, \bibinfo {author}
  {\bibfnamefont {Z.}~\bibnamefont {Tan}}, \bibinfo {author} {\bibfnamefont
  {G.}~\bibnamefont {Fan}}, \bibinfo {author} {\bibfnamefont {Z.}~\bibnamefont
  {Li}}, \ and\ \bibinfo {author} {\bibfnamefont {D.}~\bibnamefont {Zhang}},\
  }\href {\doibase 10.1021/acsnano.5b01067} {\bibfield  {journal} {\bibinfo
  {journal} {ACS Nano}\ }\textbf {\bibinfo {volume} {9}},\ \bibinfo {pages}
  {6934} (\bibinfo {year} {2015})}\BibitemShut {NoStop}%
\bibitem [{\citenamefont {Ramanathan}\ \emph {et~al.}(2008)\citenamefont
  {Ramanathan}, \citenamefont {Abdala}, \citenamefont {Stankovich},
  \citenamefont {Dikin}, \citenamefont {Herrera-Alonso}, \citenamefont {Piner},
  \citenamefont {Adamson}, \citenamefont {Schniepp}, \citenamefont {Chen},
  \citenamefont {Ruoff}, \citenamefont {Nguyen}, \citenamefont {Aksay},
  \citenamefont {Prud'Homme},\ and\ \citenamefont {Brinson}}]{RAS*08}%
  \BibitemOpen
  \bibfield  {author} {\bibinfo {author} {\bibfnamefont {T.}~\bibnamefont
  {Ramanathan}}, \bibinfo {author} {\bibfnamefont {A.~A.}\ \bibnamefont
  {Abdala}}, \bibinfo {author} {\bibfnamefont {S.}~\bibnamefont {Stankovich}},
  \bibinfo {author} {\bibfnamefont {D.~A.}\ \bibnamefont {Dikin}}, \bibinfo
  {author} {\bibfnamefont {M.}~\bibnamefont {Herrera-Alonso}}, \bibinfo
  {author} {\bibfnamefont {R.~D.}\ \bibnamefont {Piner}}, \bibinfo {author}
  {\bibfnamefont {D.~H.}\ \bibnamefont {Adamson}}, \bibinfo {author}
  {\bibfnamefont {H.~C.}\ \bibnamefont {Schniepp}}, \bibinfo {author}
  {\bibfnamefont {X.}~\bibnamefont {Chen}}, \bibinfo {author} {\bibfnamefont
  {R.~S.}\ \bibnamefont {Ruoff}}, \bibinfo {author} {\bibfnamefont {S.~T.}\
  \bibnamefont {Nguyen}}, \bibinfo {author} {\bibfnamefont {I.~A.}\
  \bibnamefont {Aksay}}, \bibinfo {author} {\bibfnamefont {R.~K.}\ \bibnamefont
  {Prud'Homme}}, \ and\ \bibinfo {author} {\bibfnamefont {L.~C.}\ \bibnamefont
  {Brinson}},\ }\href {\doibase 0.1038/nnano.2008.96} {\bibfield  {journal}
  {\bibinfo  {journal} {Nature nanotechnology}\ }\textbf {\bibinfo {volume}
  {3}},\ \bibinfo {pages} {327} (\bibinfo {year} {2008})}\BibitemShut {NoStop}%
\bibitem [{\citenamefont {Zhang}\ \emph {et~al.}(2014)\citenamefont {Zhang},
  \citenamefont {Ma}, \citenamefont {Fan}, \citenamefont {Zeng}, \citenamefont
  {Peng}, \citenamefont {Loya}, \citenamefont {Liu}, \citenamefont {Gong},
  \citenamefont {Zhang}, \citenamefont {Zhang}, \citenamefont {Ajayan},
  \citenamefont {Zhu},\ and\ \citenamefont {Lou}}]{ZMF*14}%
  \BibitemOpen
  \bibfield  {author} {\bibinfo {author} {\bibfnamefont {P.}~\bibnamefont
  {Zhang}}, \bibinfo {author} {\bibfnamefont {L.}~\bibnamefont {Ma}}, \bibinfo
  {author} {\bibfnamefont {F.}~\bibnamefont {Fan}}, \bibinfo {author}
  {\bibfnamefont {Z.}~\bibnamefont {Zeng}}, \bibinfo {author} {\bibfnamefont
  {C.}~\bibnamefont {Peng}}, \bibinfo {author} {\bibfnamefont {P.~E.}\
  \bibnamefont {Loya}}, \bibinfo {author} {\bibfnamefont {Z.}~\bibnamefont
  {Liu}}, \bibinfo {author} {\bibfnamefont {Y.}~\bibnamefont {Gong}}, \bibinfo
  {author} {\bibfnamefont {J.}~\bibnamefont {Zhang}}, \bibinfo {author}
  {\bibfnamefont {X.}~\bibnamefont {Zhang}}, \bibinfo {author} {\bibfnamefont
  {P.~M.}\ \bibnamefont {Ajayan}}, \bibinfo {author} {\bibfnamefont
  {T.}~\bibnamefont {Zhu}}, \ and\ \bibinfo {author} {\bibfnamefont
  {J.}~\bibnamefont {Lou}},\ }\href@noop {} {\bibfield  {journal} {\bibinfo
  {journal} {Nature Communications}\ }\textbf {\bibinfo {volume} {5}},\
  \bibinfo {pages} {3782 EP } (\bibinfo {year} {2014})}\BibitemShut {NoStop}%
\bibitem [{\citenamefont {Yang}\ \emph {et~al.}(2016)\citenamefont {Yang},
  \citenamefont {Wang}, \citenamefont {Lu},\ and\ \citenamefont {Hu}}]{YWLH16}%
  \BibitemOpen
  \bibfield  {author} {\bibinfo {author} {\bibfnamefont {Z.}~\bibnamefont
  {Yang}}, \bibinfo {author} {\bibfnamefont {D.}~\bibnamefont {Wang}}, \bibinfo
  {author} {\bibfnamefont {Z.}~\bibnamefont {Lu}}, \ and\ \bibinfo {author}
  {\bibfnamefont {W.}~\bibnamefont {Hu}},\ }\href {\doibase 10.1063/1.4967793}
  {\bibfield  {journal} {\bibinfo  {journal} {Applied Physics Letters}\
  }\textbf {\bibinfo {volume} {109}},\ \bibinfo {pages} {191909} (\bibinfo
  {year} {2016})}\BibitemShut {NoStop}%
\bibitem [{\citenamefont {Guo}\ \emph {et~al.}(2019)\citenamefont {Guo},
  \citenamefont {Kondoh},\ and\ \citenamefont {Han}}]{GKH19}%
  \BibitemOpen
  \bibfield  {author} {\bibinfo {author} {\bibfnamefont {Q.}~\bibnamefont
  {Guo}}, \bibinfo {author} {\bibfnamefont {K.}~\bibnamefont {Kondoh}}, \ and\
  \bibinfo {author} {\bibfnamefont {S.~M.}\ \bibnamefont {Han}},\ }\href
  {\doibase 10.1557/mrs.2018.321} {\bibfield  {journal} {\bibinfo  {journal}
  {MRS Bulletin}\ }\textbf {\bibinfo {volume} {44}},\ \bibinfo {pages} {40}
  (\bibinfo {year} {2019})}\BibitemShut {NoStop}%
\bibitem [{\citenamefont {Feng}\ \emph
  {et~al.}(2017{\natexlab{a}})\citenamefont {Feng}, \citenamefont {Song},
  \citenamefont {Xie}, \citenamefont {Wang}, \citenamefont {Liu},\ and\
  \citenamefont {Yin}}]{FSY*17}%
  \BibitemOpen
  \bibfield  {author} {\bibinfo {author} {\bibfnamefont {Q.}~\bibnamefont
  {Feng}}, \bibinfo {author} {\bibfnamefont {X.}~\bibnamefont {Song}}, \bibinfo
  {author} {\bibfnamefont {H.}~\bibnamefont {Xie}}, \bibinfo {author}
  {\bibfnamefont {H.}~\bibnamefont {Wang}}, \bibinfo {author} {\bibfnamefont
  {X.}~\bibnamefont {Liu}}, \ and\ \bibinfo {author} {\bibfnamefont
  {F.}~\bibnamefont {Yin}},\ }\href {\doibase 10.1016/j.matdes.2017.02.010}
  {\bibfield  {journal} {\bibinfo  {journal} {Materials \& Design}\ }\textbf
  {\bibinfo {volume} {120}},\ \bibinfo {pages} {193} (\bibinfo {year}
  {2017}{\natexlab{a}})}\BibitemShut {NoStop}%
\bibitem [{\citenamefont {Liu}\ \emph {et~al.}(2016)\citenamefont {Liu},
  \citenamefont {Wang}, \citenamefont {Wang},\ and\ \citenamefont
  {Wu}}]{LWWW16}%
  \BibitemOpen
  \bibfield  {author} {\bibinfo {author} {\bibfnamefont {X.}~\bibnamefont
  {Liu}}, \bibinfo {author} {\bibfnamefont {F.}~\bibnamefont {Wang}}, \bibinfo
  {author} {\bibfnamefont {W.}~\bibnamefont {Wang}}, \ and\ \bibinfo {author}
  {\bibfnamefont {H.}~\bibnamefont {Wu}},\ }\href {\doibase
  https://doi.org/10.1016/j.carbon.2016.06.071} {\bibfield  {journal} {\bibinfo
   {journal} {Carbon}\ }\textbf {\bibinfo {volume} {107}},\ \bibinfo {pages}
  {680 } (\bibinfo {year} {2016})}\BibitemShut {NoStop}%
\bibitem [{\citenamefont {Lee}\ \emph {et~al.}(2008)\citenamefont {Lee},
  \citenamefont {Wei}, \citenamefont {Kysar},\ and\ \citenamefont
  {Hone}}]{LWKH08}%
  \BibitemOpen
  \bibfield  {author} {\bibinfo {author} {\bibfnamefont {C.}~\bibnamefont
  {Lee}}, \bibinfo {author} {\bibfnamefont {X.}~\bibnamefont {Wei}}, \bibinfo
  {author} {\bibfnamefont {J.~W.}\ \bibnamefont {Kysar}}, \ and\ \bibinfo
  {author} {\bibfnamefont {J.}~\bibnamefont {Hone}},\ }\href {\doibase doi:
  10.1126/science.1157996.} {\bibfield  {journal} {\bibinfo  {journal}
  {Science}\ }\textbf {\bibinfo {volume} {321}},\ \bibinfo {pages} {385}
  (\bibinfo {year} {2008})}\BibitemShut {NoStop}%
\bibitem [{\citenamefont {Cao}\ \emph {et~al.}(2020)\citenamefont {Cao},
  \citenamefont {Feng}, \citenamefont {Han}, \citenamefont {Gao}, \citenamefont
  {Hue~Ly}, \citenamefont {Xu},\ and\ \citenamefont {Lu}}]{CFH*20}%
  \BibitemOpen
  \bibfield  {author} {\bibinfo {author} {\bibfnamefont {K.}~\bibnamefont
  {Cao}}, \bibinfo {author} {\bibfnamefont {S.}~\bibnamefont {Feng}}, \bibinfo
  {author} {\bibfnamefont {Y.}~\bibnamefont {Han}}, \bibinfo {author}
  {\bibfnamefont {L.}~\bibnamefont {Gao}}, \bibinfo {author} {\bibfnamefont
  {T.}~\bibnamefont {Hue~Ly}}, \bibinfo {author} {\bibfnamefont
  {Z.}~\bibnamefont {Xu}}, \ and\ \bibinfo {author} {\bibfnamefont
  {Y.}~\bibnamefont {Lu}},\ }\href {\doibase 10.1038/s41467-019-14130-0}
  {\bibfield  {journal} {\bibinfo  {journal} {Nature Communications}\ }\textbf
  {\bibinfo {volume} {11}},\ \bibinfo {pages} {284} (\bibinfo {year}
  {2020})}\BibitemShut {NoStop}%
\bibitem [{\citenamefont {Feng}\ \emph
  {et~al.}(2017{\natexlab{b}})\citenamefont {Feng}, \citenamefont {Guo},
  \citenamefont {Li}, \citenamefont {Fan}, \citenamefont {Li}, \citenamefont
  {Xiong}, \citenamefont {Su}, \citenamefont {Tan}, \citenamefont {Zhang},\
  and\ \citenamefont {Zhang}}]{FGL*17}%
  \BibitemOpen
  \bibfield  {author} {\bibinfo {author} {\bibfnamefont {S.}~\bibnamefont
  {Feng}}, \bibinfo {author} {\bibfnamefont {Q.}~\bibnamefont {Guo}}, \bibinfo
  {author} {\bibfnamefont {Z.}~\bibnamefont {Li}}, \bibinfo {author}
  {\bibfnamefont {G.}~\bibnamefont {Fan}}, \bibinfo {author} {\bibfnamefont
  {Z.}~\bibnamefont {Li}}, \bibinfo {author} {\bibfnamefont {D.-B.}\
  \bibnamefont {Xiong}}, \bibinfo {author} {\bibfnamefont {Y.}~\bibnamefont
  {Su}}, \bibinfo {author} {\bibfnamefont {Z.}~\bibnamefont {Tan}}, \bibinfo
  {author} {\bibfnamefont {J.}~\bibnamefont {Zhang}}, \ and\ \bibinfo {author}
  {\bibfnamefont {D.}~\bibnamefont {Zhang}},\ }\href {\doibase
  http://dx.doi.org/10.1016/j.actamat.2016.11.043} {\bibfield  {journal}
  {\bibinfo  {journal} {Acta Mater.}\ }\textbf {\bibinfo {volume} {125}},\
  \bibinfo {pages} {98 } (\bibinfo {year} {2017}{\natexlab{b}})}\BibitemShut
  {NoStop}%
\bibitem [{\citenamefont {Hwang}\ \emph {et~al.}(2013)\citenamefont {Hwang},
  \citenamefont {Yoon}, \citenamefont {Jin}, \citenamefont {Lee}, \citenamefont
  {Kim}, \citenamefont {Hong},\ and\ \citenamefont {Jeon}}]{HYJ*13}%
  \BibitemOpen
  \bibfield  {author} {\bibinfo {author} {\bibfnamefont {J.}~\bibnamefont
  {Hwang}}, \bibinfo {author} {\bibfnamefont {T.}~\bibnamefont {Yoon}},
  \bibinfo {author} {\bibfnamefont {S.~H.}\ \bibnamefont {Jin}}, \bibinfo
  {author} {\bibfnamefont {J.}~\bibnamefont {Lee}}, \bibinfo {author}
  {\bibfnamefont {T.-S.}\ \bibnamefont {Kim}}, \bibinfo {author} {\bibfnamefont
  {S.~H.}\ \bibnamefont {Hong}}, \ and\ \bibinfo {author} {\bibfnamefont
  {S.}~\bibnamefont {Jeon}},\ }\href {\doibase
  https://doi.org/10.1002/adma.201302495} {\bibfield  {journal} {\bibinfo
  {journal} {Advanced Materials}\ }\textbf {\bibinfo {volume} {25}},\ \bibinfo
  {pages} {6724} (\bibinfo {year} {2013})}\BibitemShut {NoStop}%
\bibitem [{\citenamefont {Khan}\ \emph {et~al.}(2015)\citenamefont {Khan},
  \citenamefont {Tahir}, \citenamefont {Adil}, \citenamefont {Khan},
  \citenamefont {Siddiqui}, \citenamefont {Al-warthan},\ and\ \citenamefont
  {Tremel}}]{KTA*15}%
  \BibitemOpen
  \bibfield  {author} {\bibinfo {author} {\bibfnamefont {M.}~\bibnamefont
  {Khan}}, \bibinfo {author} {\bibfnamefont {M.~N.}\ \bibnamefont {Tahir}},
  \bibinfo {author} {\bibfnamefont {S.~F.}\ \bibnamefont {Adil}}, \bibinfo
  {author} {\bibfnamefont {H.~U.}\ \bibnamefont {Khan}}, \bibinfo {author}
  {\bibfnamefont {M.~R.~H.}\ \bibnamefont {Siddiqui}}, \bibinfo {author}
  {\bibfnamefont {A.~A.}\ \bibnamefont {Al-warthan}}, \ and\ \bibinfo {author}
  {\bibfnamefont {W.}~\bibnamefont {Tremel}},\ }\href {\doibase
  10.1039/C5TA02240A} {\bibfield  {journal} {\bibinfo  {journal} {J. Mater.
  Chem. A}\ }\textbf {\bibinfo {volume} {3}},\ \bibinfo {pages} {18753}
  (\bibinfo {year} {2015})}\BibitemShut {NoStop}%
\bibitem [{\citenamefont {Hwang}\ \emph {et~al.}(2017)\citenamefont {Hwang},
  \citenamefont {Kim}, \citenamefont {Kim}, \citenamefont {Lee}, \citenamefont
  {Lim}, \citenamefont {Kim}, \citenamefont {Oh}, \citenamefont {Ryu},\ and\
  \citenamefont {Han}}]{HKK*17}%
  \BibitemOpen
  \bibfield  {author} {\bibinfo {author} {\bibfnamefont {B.}~\bibnamefont
  {Hwang}}, \bibinfo {author} {\bibfnamefont {W.}~\bibnamefont {Kim}}, \bibinfo
  {author} {\bibfnamefont {J.}~\bibnamefont {Kim}}, \bibinfo {author}
  {\bibfnamefont {S.}~\bibnamefont {Lee}}, \bibinfo {author} {\bibfnamefont
  {S.}~\bibnamefont {Lim}}, \bibinfo {author} {\bibfnamefont {S.}~\bibnamefont
  {Kim}}, \bibinfo {author} {\bibfnamefont {S.~H.}\ \bibnamefont {Oh}},
  \bibinfo {author} {\bibfnamefont {S.}~\bibnamefont {Ryu}}, \ and\ \bibinfo
  {author} {\bibfnamefont {S.~M.}\ \bibnamefont {Han}},\ }\href {\doibase
  10.1021/acs.nanolett.7b01431} {\bibfield  {journal} {\bibinfo  {journal}
  {Nano Letters}\ }\textbf {\bibinfo {volume} {17}},\ \bibinfo {pages} {4740}
  (\bibinfo {year} {2017})}\BibitemShut {NoStop}%
\bibitem [{\citenamefont {Li}\ \emph {et~al.}(2015)\citenamefont {Li},
  \citenamefont {Guo}, \citenamefont {Li}, \citenamefont {Fan}, \citenamefont
  {Xiong}, \citenamefont {Su}, \citenamefont {Zhang},\ and\ \citenamefont
  {Zhang}}]{LGL*15}%
  \BibitemOpen
  \bibfield  {author} {\bibinfo {author} {\bibfnamefont {Z.}~\bibnamefont
  {Li}}, \bibinfo {author} {\bibfnamefont {Q.}~\bibnamefont {Guo}}, \bibinfo
  {author} {\bibfnamefont {Z.}~\bibnamefont {Li}}, \bibinfo {author}
  {\bibfnamefont {G.}~\bibnamefont {Fan}}, \bibinfo {author} {\bibfnamefont
  {D.-B.}\ \bibnamefont {Xiong}}, \bibinfo {author} {\bibfnamefont
  {Y.}~\bibnamefont {Su}}, \bibinfo {author} {\bibfnamefont {J.}~\bibnamefont
  {Zhang}}, \ and\ \bibinfo {author} {\bibfnamefont {D.}~\bibnamefont
  {Zhang}},\ }\href {\doibase 10.1021/acs.nanolett.5b03492} {\bibfield
  {journal} {\bibinfo  {journal} {Nano Lett.}\ }\textbf {\bibinfo {volume}
  {15}},\ \bibinfo {pages} {8077} (\bibinfo {year} {2015})}\BibitemShut
  {NoStop}%
\bibitem [{\citenamefont {Kim}\ \emph {et~al.}(2013)\citenamefont {Kim},
  \citenamefont {Lee}, \citenamefont {Yeom}, \citenamefont {Shin},
  \citenamefont {Kim}, \citenamefont {Cui}, \citenamefont {Kysar},
  \citenamefont {Hone}, \citenamefont {Jung}, \citenamefont {Jeon},\ and\
  \citenamefont {Han}}]{KLY*13}%
  \BibitemOpen
  \bibfield  {author} {\bibinfo {author} {\bibfnamefont {Y.}~\bibnamefont
  {Kim}}, \bibinfo {author} {\bibfnamefont {J.}~\bibnamefont {Lee}}, \bibinfo
  {author} {\bibfnamefont {M.~S.}\ \bibnamefont {Yeom}}, \bibinfo {author}
  {\bibfnamefont {J.~W.}\ \bibnamefont {Shin}}, \bibinfo {author}
  {\bibfnamefont {H.}~\bibnamefont {Kim}}, \bibinfo {author} {\bibfnamefont
  {Y.}~\bibnamefont {Cui}}, \bibinfo {author} {\bibfnamefont {J.~W.}\
  \bibnamefont {Kysar}}, \bibinfo {author} {\bibfnamefont {J.}~\bibnamefont
  {Hone}}, \bibinfo {author} {\bibfnamefont {Y.}~\bibnamefont {Jung}}, \bibinfo
  {author} {\bibfnamefont {S.}~\bibnamefont {Jeon}}, \ and\ \bibinfo {author}
  {\bibfnamefont {S.~M.}\ \bibnamefont {Han}},\ }\href {\doibase
  10.1038/ncomms3114} {\bibfield  {journal} {\bibinfo  {journal} {Nat.
  Commun.}\ }\textbf {\bibinfo {volume} {4}},\ \bibinfo {pages} {2114}
  (\bibinfo {year} {2013})}\BibitemShut {NoStop}%
\bibitem [{\citenamefont {Zhou}\ and\ \citenamefont {Chen}(2016)}]{ZC16}%
  \BibitemOpen
  \bibfield  {author} {\bibinfo {author} {\bibfnamefont {X.}~\bibnamefont
  {Zhou}}\ and\ \bibinfo {author} {\bibfnamefont {C.}~\bibnamefont {Chen}},\
  }\href {\doibase https://doi.org/10.1016/j.ijplas.2016.01.003} {\bibfield
  {journal} {\bibinfo  {journal} {International Journal of Plasticity}\
  }\textbf {\bibinfo {volume} {80}},\ \bibinfo {pages} {75 } (\bibinfo {year}
  {2016})}\BibitemShut {NoStop}%
\bibitem [{\citenamefont {Sterwerf}\ \emph {et~al.}(2017)\citenamefont
  {Sterwerf}, \citenamefont {Kaub}, \citenamefont {Deng}, \citenamefont
  {Thompson},\ and\ \citenamefont {Li}}]{SKD*17}%
  \BibitemOpen
  \bibfield  {author} {\bibinfo {author} {\bibfnamefont {C.}~\bibnamefont
  {Sterwerf}}, \bibinfo {author} {\bibfnamefont {T.}~\bibnamefont {Kaub}},
  \bibinfo {author} {\bibfnamefont {C.}~\bibnamefont {Deng}}, \bibinfo {author}
  {\bibfnamefont {G.~B.}\ \bibnamefont {Thompson}}, \ and\ \bibinfo {author}
  {\bibfnamefont {L.}~\bibnamefont {Li}},\ }\href {\doibase
  https://doi.org/10.1016/j.tsf.2017.02.035} {\bibfield  {journal} {\bibinfo
  {journal} {Thin Solid Films}\ }\textbf {\bibinfo {volume} {626}},\ \bibinfo
  {pages} {184 } (\bibinfo {year} {2017})}\BibitemShut {NoStop}%
\bibitem [{\citenamefont {Song}\ \emph {et~al.}(2017)\citenamefont {Song},
  \citenamefont {Xu}, \citenamefont {Zhang}, \citenamefont {Li}, \citenamefont
  {Wang},\ and\ \citenamefont {Li}}]{SXZ*17}%
  \BibitemOpen
  \bibfield  {author} {\bibinfo {author} {\bibfnamefont {H.~Y.}\ \bibnamefont
  {Song}}, \bibinfo {author} {\bibfnamefont {J.~J.}\ \bibnamefont {Xu}},
  \bibinfo {author} {\bibfnamefont {Y.~G.}\ \bibnamefont {Zhang}}, \bibinfo
  {author} {\bibfnamefont {S.}~\bibnamefont {Li}}, \bibinfo {author}
  {\bibfnamefont {D.~H.}\ \bibnamefont {Wang}}, \ and\ \bibinfo {author}
  {\bibfnamefont {Y.~L.}\ \bibnamefont {Li}},\ }\href {\doibase
  https://doi.org/10.1016/j.matdes.2017.04.077} {\bibfield  {journal} {\bibinfo
   {journal} {Materials \& Design}\ }\textbf {\bibinfo {volume} {127}},\
  \bibinfo {pages} {173 } (\bibinfo {year} {2017})}\BibitemShut {NoStop}%
\bibitem [{\citenamefont {Avila}\ \emph {et~al.}(2021)\citenamefont {Avila},
  \citenamefont {Vardanyan}, \citenamefont {K\"uchemann},\ and\ \citenamefont
  {Urbassek}}]{AVKU21}%
  \BibitemOpen
  \bibfield  {author} {\bibinfo {author} {\bibfnamefont {K.~E.}\ \bibnamefont
  {Avila}}, \bibinfo {author} {\bibfnamefont {V.~H.}\ \bibnamefont
  {Vardanyan}}, \bibinfo {author} {\bibfnamefont {S.}~\bibnamefont
  {K\"uchemann}}, \ and\ \bibinfo {author} {\bibfnamefont {H.~M.}\ \bibnamefont
  {Urbassek}},\ }\href {\doibase https://doi.org/10.1016/j.apsusc.2021.149285}
  {\bibfield  {journal} {\bibinfo  {journal} {Applied Surface Science}\
  }\textbf {\bibinfo {volume} {551}},\ \bibinfo {pages} {149285} (\bibinfo
  {year} {2021})}\BibitemShut {NoStop}%
\bibitem [{\citenamefont {Wang}\ and\ \citenamefont {Misra}(2011)}]{WM11}%
  \BibitemOpen
  \bibfield  {author} {\bibinfo {author} {\bibfnamefont {J.}~\bibnamefont
  {Wang}}\ and\ \bibinfo {author} {\bibfnamefont {A.}~\bibnamefont {Misra}},\
  }\href {\doibase 10.1016/j.cossms.2010.09.002} {\bibfield  {journal}
  {\bibinfo  {journal} {Current Opinion in Solid State and Materials Science}\
  }\textbf {\bibinfo {volume} {15}},\ \bibinfo {pages} {20} (\bibinfo {year}
  {2011})}\BibitemShut {NoStop}%
\bibitem [{\citenamefont {Shabib}\ and\ \citenamefont {Miller}(2009)}]{SM09}%
  \BibitemOpen
  \bibfield  {author} {\bibinfo {author} {\bibfnamefont {I.}~\bibnamefont
  {Shabib}}\ and\ \bibinfo {author} {\bibfnamefont {R.~E.}\ \bibnamefont
  {Miller}},\ }\href {\doibase https://doi.org/10.1016/j.actamat.2009.05.028}
  {\bibfield  {journal} {\bibinfo  {journal} {Acta Materialia}\ }\textbf
  {\bibinfo {volume} {57}},\ \bibinfo {pages} {4364 } (\bibinfo {year}
  {2009})}\BibitemShut {NoStop}%
\bibitem [{\citenamefont {Schi{\o}tz}\ \emph {et~al.}(1998)\citenamefont
  {Schi{\o}tz}, \citenamefont {Di~Tolla},\ and\ \citenamefont
  {Jacobsen}}]{STJ98}%
  \BibitemOpen
  \bibfield  {author} {\bibinfo {author} {\bibfnamefont {J.}~\bibnamefont
  {Schi{\o}tz}}, \bibinfo {author} {\bibfnamefont {F.~D.}\ \bibnamefont
  {Di~Tolla}}, \ and\ \bibinfo {author} {\bibfnamefont {K.~W.}\ \bibnamefont
  {Jacobsen}},\ }\href {\doibase 10.1038/35328} {\bibfield  {journal} {\bibinfo
   {journal} {Nature}\ }\textbf {\bibinfo {volume} {391}},\ \bibinfo {pages}
  {561} (\bibinfo {year} {1998})}\BibitemShut {NoStop}%
\bibitem [{\citenamefont {Van~Swygenhoven}\ and\ \citenamefont
  {Weertman}(2006)}]{SW06}%
  \BibitemOpen
  \bibfield  {author} {\bibinfo {author} {\bibfnamefont {H.}~\bibnamefont
  {Van~Swygenhoven}}\ and\ \bibinfo {author} {\bibfnamefont {J.~R.}\
  \bibnamefont {Weertman}},\ }\href {\doibase
  https://doi.org/10.1016/S1369-7021(06)71494-8} {\bibfield  {journal}
  {\bibinfo  {journal} {Materials Today}\ }\textbf {\bibinfo {volume} {9}},\
  \bibinfo {pages} {24} (\bibinfo {year} {2006})}\BibitemShut {NoStop}%
\bibitem [{\citenamefont {Zhou}\ \emph {et~al.}(2020)\citenamefont {Zhou},
  \citenamefont {Feng}, \citenamefont {Zhu}, \citenamefont {Xu}, \citenamefont
  {Miyagi}, \citenamefont {Dong}, \citenamefont {Sheng}, \citenamefont {Wang},
  \citenamefont {Li}, \citenamefont {Ma}, \citenamefont {Zhang}, \citenamefont
  {Yan}, \citenamefont {Tamura}, \citenamefont {Kunz}, \citenamefont {Lutker},
  \citenamefont {Huang}, \citenamefont {Hughes}, \citenamefont {Huang},\ and\
  \citenamefont {Chen}}]{ZFZ*20}%
  \BibitemOpen
  \bibfield  {author} {\bibinfo {author} {\bibfnamefont {X.}~\bibnamefont
  {Zhou}}, \bibinfo {author} {\bibfnamefont {Z.}~\bibnamefont {Feng}}, \bibinfo
  {author} {\bibfnamefont {L.}~\bibnamefont {Zhu}}, \bibinfo {author}
  {\bibfnamefont {J.}~\bibnamefont {Xu}}, \bibinfo {author} {\bibfnamefont
  {L.}~\bibnamefont {Miyagi}}, \bibinfo {author} {\bibfnamefont
  {H.}~\bibnamefont {Dong}}, \bibinfo {author} {\bibfnamefont {H.}~\bibnamefont
  {Sheng}}, \bibinfo {author} {\bibfnamefont {Y.}~\bibnamefont {Wang}},
  \bibinfo {author} {\bibfnamefont {Q.}~\bibnamefont {Li}}, \bibinfo {author}
  {\bibfnamefont {Y.}~\bibnamefont {Ma}}, \bibinfo {author} {\bibfnamefont
  {H.}~\bibnamefont {Zhang}}, \bibinfo {author} {\bibfnamefont
  {J.}~\bibnamefont {Yan}}, \bibinfo {author} {\bibfnamefont {N.}~\bibnamefont
  {Tamura}}, \bibinfo {author} {\bibfnamefont {M.}~\bibnamefont {Kunz}},
  \bibinfo {author} {\bibfnamefont {K.}~\bibnamefont {Lutker}}, \bibinfo
  {author} {\bibfnamefont {T.}~\bibnamefont {Huang}}, \bibinfo {author}
  {\bibfnamefont {D.~A.}\ \bibnamefont {Hughes}}, \bibinfo {author}
  {\bibfnamefont {X.}~\bibnamefont {Huang}}, \ and\ \bibinfo {author}
  {\bibfnamefont {B.}~\bibnamefont {Chen}},\ }\href {\doibase
  10.1038/s41586-020-2036-z} {\bibfield  {journal} {\bibinfo  {journal}
  {Nature}\ }\textbf {\bibinfo {volume} {579}},\ \bibinfo {pages} {67}
  (\bibinfo {year} {2020})}\BibitemShut {NoStop}%
\bibitem [{\citenamefont {Chang}\ \emph {et~al.}(2013)\citenamefont {Chang},
  \citenamefont {Nair},\ and\ \citenamefont {Buehler}}]{CNB13}%
  \BibitemOpen
  \bibfield  {author} {\bibinfo {author} {\bibfnamefont {S.-W.}\ \bibnamefont
  {Chang}}, \bibinfo {author} {\bibfnamefont {A.~K.}\ \bibnamefont {Nair}}, \
  and\ \bibinfo {author} {\bibfnamefont {M.~J.}\ \bibnamefont {Buehler}},\
  }\href {\doibase 10.1080/09500839.2012.759293} {\bibfield  {journal}
  {\bibinfo  {journal} {Philosophical Magazine Letters}\ }\textbf {\bibinfo
  {volume} {93}},\ \bibinfo {pages} {196} (\bibinfo {year} {2013})}\BibitemShut
  {NoStop}%
\bibitem [{\citenamefont {Muller}\ \emph {et~al.}(2018)\citenamefont {Muller},
  \citenamefont {Santhapuram},\ and\ \citenamefont {Nair}}]{MSN18}%
  \BibitemOpen
  \bibfield  {author} {\bibinfo {author} {\bibfnamefont {S.~E.}\ \bibnamefont
  {Muller}}, \bibinfo {author} {\bibfnamefont {R.~R.}\ \bibnamefont
  {Santhapuram}}, \ and\ \bibinfo {author} {\bibfnamefont {A.~K.}\ \bibnamefont
  {Nair}},\ }\href {\doibase https://doi.org/10.1016/j.commatsci.2018.06.013}
  {\bibfield  {journal} {\bibinfo  {journal} {Computational Materials Science}\
  }\textbf {\bibinfo {volume} {152}},\ \bibinfo {pages} {341 } (\bibinfo {year}
  {2018})}\BibitemShut {NoStop}%
\bibitem [{\citenamefont {~}\ \emph {et~al.}(2017)\citenamefont {~},
  \citenamefont {Yari~Boroujeni},\ and\ \citenamefont {Mirzaeifar}}]{YBM17}%
  \BibitemOpen
  \bibfield  {author} {\bibinfo {author} {\bibfnamefont {F.}~\bibnamefont {~}},
  \bibinfo {author} {\bibfnamefont {A.}~\bibnamefont {Yari~Boroujeni}}, \ and\
  \bibinfo {author} {\bibfnamefont {R.}~\bibnamefont {Mirzaeifar}},\ }\href
  {\doibase 10.1103/PhysRevMaterials.1.076001} {\bibfield  {journal} {\bibinfo
  {journal} {Phys. Rev. Materials}\ }\textbf {\bibinfo {volume} {1}},\ \bibinfo
  {pages} {076001} (\bibinfo {year} {2017})}\BibitemShut {NoStop}%
\bibitem [{\citenamefont {Vardanyan}\ and\ \citenamefont
  {Urbassek}(2019)}]{VU19}%
  \BibitemOpen
  \bibfield  {author} {\bibinfo {author} {\bibfnamefont {V.~H.}\ \bibnamefont
  {Vardanyan}}\ and\ \bibinfo {author} {\bibfnamefont {H.~M.}\ \bibnamefont
  {Urbassek}},\ }\href {\doibase
  https://doi.org/10.1016/j.commatsci.2019.109158} {\bibfield  {journal}
  {\bibinfo  {journal} {Computational Materials Science}\ }\textbf {\bibinfo
  {volume} {170}},\ \bibinfo {pages} {109158} (\bibinfo {year}
  {2019})}\BibitemShut {NoStop}%
\bibitem [{\citenamefont {Vardanyan}\ and\ \citenamefont
  {Urbassek}(2020)}]{VU20}%
  \BibitemOpen
  \bibfield  {author} {\bibinfo {author} {\bibfnamefont {V.~H.}\ \bibnamefont
  {Vardanyan}}\ and\ \bibinfo {author} {\bibfnamefont {H.~M.}\ \bibnamefont
  {Urbassek}},\ }\href {\doibase 10.3390/ma13071683} {\bibfield  {journal}
  {\bibinfo  {journal} {Materials}\ }\textbf {\bibinfo {volume} {13}},\
  \bibinfo {pages} {1683} (\bibinfo {year} {2020})}\BibitemShut {NoStop}%
\bibitem [{\citenamefont {Shuang}\ and\ \citenamefont {Aifantis}(2021)}]{SA21}%
  \BibitemOpen
  \bibfield  {author} {\bibinfo {author} {\bibfnamefont {F.}~\bibnamefont
  {Shuang}}\ and\ \bibinfo {author} {\bibfnamefont {K.~E.}\ \bibnamefont
  {Aifantis}},\ }\href {\doibase https://doi.org/10.1016/j.carbon.2020.09.043}
  {\bibfield  {journal} {\bibinfo  {journal} {Carbon}\ }\textbf {\bibinfo
  {volume} {172}},\ \bibinfo {pages} {50} (\bibinfo {year} {2021})}\BibitemShut
  {NoStop}%
\bibitem [{\citenamefont {Shuang}\ \emph {et~al.}(2021)\citenamefont {Shuang},
  \citenamefont {Dai},\ and\ \citenamefont {Aifantis}}]{SDA21}%
  \BibitemOpen
  \bibfield  {author} {\bibinfo {author} {\bibfnamefont {F.}~\bibnamefont
  {Shuang}}, \bibinfo {author} {\bibfnamefont {Z.}~\bibnamefont {Dai}}, \ and\
  \bibinfo {author} {\bibfnamefont {K.~E.}\ \bibnamefont {Aifantis}},\ }\href
  {\doibase 10.1021/acsami.1c05129} {\bibfield  {journal} {\bibinfo  {journal}
  {ACS Applied Materials \& Interfaces}\ }\textbf {\bibinfo {volume} {13}},\
  \bibinfo {pages} {26610} (\bibinfo {year} {2021})}\BibitemShut {NoStop}%
\bibitem [{\citenamefont {Zhang}\ \emph
  {et~al.}(2020{\natexlab{a}})\citenamefont {Zhang}, \citenamefont {Huang},\
  and\ \citenamefont {Wang}}]{ZHW20}%
  \BibitemOpen
  \bibfield  {author} {\bibinfo {author} {\bibfnamefont {S.}~\bibnamefont
  {Zhang}}, \bibinfo {author} {\bibfnamefont {P.}~\bibnamefont {Huang}}, \ and\
  \bibinfo {author} {\bibfnamefont {F.}~\bibnamefont {Wang}},\ }\href {\doibase
  https://doi.org/10.1016/j.matdes.2020.108555} {\bibfield  {journal} {\bibinfo
   {journal} {Materials \& Design}\ }\textbf {\bibinfo {volume} {190}},\
  \bibinfo {pages} {108555} (\bibinfo {year} {2020}{\natexlab{a}})}\BibitemShut
  {NoStop}%
\bibitem [{\citenamefont {Peng}\ and\ \citenamefont {Sun}(2020)}]{PS20}%
  \BibitemOpen
  \bibfield  {author} {\bibinfo {author} {\bibfnamefont {W.}~\bibnamefont
  {Peng}}\ and\ \bibinfo {author} {\bibfnamefont {K.}~\bibnamefont {Sun}},\
  }\href {\doibase https://doi.org/10.1016/j.mechmat.2019.103270} {\bibfield
  {journal} {\bibinfo  {journal} {Mechanics of Materials}\ }\textbf {\bibinfo
  {volume} {141}},\ \bibinfo {pages} {103270} (\bibinfo {year}
  {2020})}\BibitemShut {NoStop}%
\bibitem [{\citenamefont {Rezaei}(2018)}]{Rez18}%
  \BibitemOpen
  \bibfield  {author} {\bibinfo {author} {\bibfnamefont {R.}~\bibnamefont
  {Rezaei}},\ }\href {\doibase https://doi.org/10.1016/j.commatsci.2018.05.004}
  {\bibfield  {journal} {\bibinfo  {journal} {Computational Materials Science}\
  }\textbf {\bibinfo {volume} {151}},\ \bibinfo {pages} {181} (\bibinfo {year}
  {2018})}\BibitemShut {NoStop}%
\bibitem [{\citenamefont {Ma}\ \emph {et~al.}(2020)\citenamefont {Ma},
  \citenamefont {Zhang}, \citenamefont {Xu}, \citenamefont {Liu},\ and\
  \citenamefont {Luo}}]{MZX*20}%
  \BibitemOpen
  \bibfield  {author} {\bibinfo {author} {\bibfnamefont {Y.}~\bibnamefont
  {Ma}}, \bibinfo {author} {\bibfnamefont {S.}~\bibnamefont {Zhang}}, \bibinfo
  {author} {\bibfnamefont {Y.}~\bibnamefont {Xu}}, \bibinfo {author}
  {\bibfnamefont {X.}~\bibnamefont {Liu}}, \ and\ \bibinfo {author}
  {\bibfnamefont {S.-N.}\ \bibnamefont {Luo}},\ }\href {\doibase
  10.1039/C9CP06830A} {\bibfield  {journal} {\bibinfo  {journal} {Phys. Chem.
  Chem. Phys.}\ }\textbf {\bibinfo {volume} {22}},\ \bibinfo {pages} {4741}
  (\bibinfo {year} {2020})}\BibitemShut {NoStop}%
\bibitem [{\citenamefont {Shuang}\ and\ \citenamefont {Aifantis}(2020)}]{SA20}%
  \BibitemOpen
  \bibfield  {author} {\bibinfo {author} {\bibfnamefont {F.}~\bibnamefont
  {Shuang}}\ and\ \bibinfo {author} {\bibfnamefont {K.~E.}\ \bibnamefont
  {Aifantis}},\ }\href {\doibase
  https://doi.org/10.1016/j.scriptamat.2020.02.014} {\bibfield  {journal}
  {\bibinfo  {journal} {Scripta Materialia}\ }\textbf {\bibinfo {volume}
  {181}},\ \bibinfo {pages} {70 } (\bibinfo {year} {2020})}\BibitemShut
  {NoStop}%
\bibitem [{\citenamefont {Zhang}\ \emph
  {et~al.}(2019{\natexlab{a}})\citenamefont {Zhang}, \citenamefont {Lu},
  \citenamefont {Pei}, \citenamefont {Li},\ and\ \citenamefont
  {Wang}}]{ZLP*19}%
  \BibitemOpen
  \bibfield  {author} {\bibinfo {author} {\bibfnamefont {C.}~\bibnamefont
  {Zhang}}, \bibinfo {author} {\bibfnamefont {C.}~\bibnamefont {Lu}}, \bibinfo
  {author} {\bibfnamefont {L.}~\bibnamefont {Pei}}, \bibinfo {author}
  {\bibfnamefont {J.}~\bibnamefont {Li}}, \ and\ \bibinfo {author}
  {\bibfnamefont {R.}~\bibnamefont {Wang}},\ }\href {\doibase
  https://doi.org/10.1016/j.compositesb.2019.107610} {\bibfield  {journal}
  {\bibinfo  {journal} {Composites Part B: Engineering}\ ,\ \bibinfo {pages}
  {107610}} (\bibinfo {year} {2019}{\natexlab{a}})}\BibitemShut {NoStop}%
\bibitem [{\citenamefont {Rezaei}\ \emph {et~al.}(2016)\citenamefont {Rezaei},
  \citenamefont {Deng}, \citenamefont {Tavakoli-Anbaran},\ and\ \citenamefont
  {Shariati}}]{RDTS16}%
  \BibitemOpen
  \bibfield  {author} {\bibinfo {author} {\bibfnamefont {R.}~\bibnamefont
  {Rezaei}}, \bibinfo {author} {\bibfnamefont {C.}~\bibnamefont {Deng}},
  \bibinfo {author} {\bibfnamefont {H.}~\bibnamefont {Tavakoli-Anbaran}}, \
  and\ \bibinfo {author} {\bibfnamefont {M.}~\bibnamefont {Shariati}},\ }\href
  {\doibase 10.1080/09500839.2016.1216195} {\bibfield  {journal} {\bibinfo
  {journal} {Philosophical Magazine Letters}\ }\textbf {\bibinfo {volume}
  {96}},\ \bibinfo {pages} {322} (\bibinfo {year} {2016})}\BibitemShut
  {NoStop}%
\bibitem [{\citenamefont {Zhang}\ \emph
  {et~al.}(2019{\natexlab{b}})\citenamefont {Zhang}, \citenamefont {Lu},
  \citenamefont {Pei}, \citenamefont {Li}, \citenamefont {Wang},\ and\
  \citenamefont {Tieu}}]{ZLP*19a}%
  \BibitemOpen
  \bibfield  {author} {\bibinfo {author} {\bibfnamefont {C.}~\bibnamefont
  {Zhang}}, \bibinfo {author} {\bibfnamefont {C.}~\bibnamefont {Lu}}, \bibinfo
  {author} {\bibfnamefont {L.}~\bibnamefont {Pei}}, \bibinfo {author}
  {\bibfnamefont {J.}~\bibnamefont {Li}}, \bibinfo {author} {\bibfnamefont
  {R.}~\bibnamefont {Wang}}, \ and\ \bibinfo {author} {\bibfnamefont
  {K.}~\bibnamefont {Tieu}},\ }\href {\doibase
  https://doi.org/10.1016/j.carbon.2018.10.097} {\bibfield  {journal} {\bibinfo
   {journal} {Carbon}\ }\textbf {\bibinfo {volume} {143}},\ \bibinfo {pages}
  {125} (\bibinfo {year} {2019}{\natexlab{b}})}\BibitemShut {NoStop}%
\bibitem [{\citenamefont {Xia}\ \emph {et~al.}(2021)\citenamefont {Xia},
  \citenamefont {Du}, \citenamefont {Zhang}, \citenamefont {Li},\ and\
  \citenamefont {Weng}}]{XDZ*21}%
  \BibitemOpen
  \bibfield  {author} {\bibinfo {author} {\bibfnamefont {X.}~\bibnamefont
  {Xia}}, \bibinfo {author} {\bibfnamefont {Z.}~\bibnamefont {Du}}, \bibinfo
  {author} {\bibfnamefont {J.}~\bibnamefont {Zhang}}, \bibinfo {author}
  {\bibfnamefont {J.}~\bibnamefont {Li}}, \ and\ \bibinfo {author}
  {\bibfnamefont {G.~J.}\ \bibnamefont {Weng}},\ }\href {\doibase
  https://doi.org/10.1016/j.ijengsci.2021.103476} {\bibfield  {journal}
  {\bibinfo  {journal} {International Journal of Engineering Science}\ }\textbf
  {\bibinfo {volume} {162}},\ \bibinfo {pages} {103476} (\bibinfo {year}
  {2021})}\BibitemShut {NoStop}%
\bibitem [{\citenamefont {Wei}\ \emph {et~al.}(2021)\citenamefont {Wei},
  \citenamefont {Ye}, \citenamefont {Hong}, \citenamefont {Yao}, \citenamefont
  {Xia}, \citenamefont {Mao}, \citenamefont {Wang}, \citenamefont {Zhao},\ and\
  \citenamefont {Tang}}]{WYH*21}%
  \BibitemOpen
  \bibfield  {author} {\bibinfo {author} {\bibfnamefont {C.}~\bibnamefont
  {Wei}}, \bibinfo {author} {\bibfnamefont {N.}~\bibnamefont {Ye}}, \bibinfo
  {author} {\bibfnamefont {L.}~\bibnamefont {Hong}}, \bibinfo {author}
  {\bibfnamefont {J.}~\bibnamefont {Yao}}, \bibinfo {author} {\bibfnamefont
  {W.}~\bibnamefont {Xia}}, \bibinfo {author} {\bibfnamefont {J.}~\bibnamefont
  {Mao}}, \bibinfo {author} {\bibfnamefont {Y.}~\bibnamefont {Wang}}, \bibinfo
  {author} {\bibfnamefont {Y.}~\bibnamefont {Zhao}}, \ and\ \bibinfo {author}
  {\bibfnamefont {J.}~\bibnamefont {Tang}},\ }\href {\doibase
  10.1021/acsami.1c01519} {\bibfield  {journal} {\bibinfo  {journal} {ACS
  Applied Materials \& Interfaces}\ }\textbf {\bibinfo {volume} {13}},\
  \bibinfo {pages} {21714} (\bibinfo {year} {2021})}\BibitemShut {NoStop}%
\bibitem [{\citenamefont {Han}\ \emph {et~al.}(2021)\citenamefont {Han},
  \citenamefont {Song},\ and\ \citenamefont {An}}]{HSA21}%
  \BibitemOpen
  \bibfield  {author} {\bibinfo {author} {\bibfnamefont {R.~Q.}\ \bibnamefont
  {Han}}, \bibinfo {author} {\bibfnamefont {H.~Y.}\ \bibnamefont {Song}}, \
  and\ \bibinfo {author} {\bibfnamefont {M.~R.}\ \bibnamefont {An}},\ }\href
  {\doibase https://doi.org/10.1016/j.commatsci.2021.110604} {\bibfield
  {journal} {\bibinfo  {journal} {Computational Materials Science}\ }\textbf
  {\bibinfo {volume} {197}},\ \bibinfo {pages} {110604} (\bibinfo {year}
  {2021})}\BibitemShut {NoStop}%
\bibitem [{\citenamefont {Rong}\ \emph {et~al.}(2018)\citenamefont {Rong},
  \citenamefont {He}, \citenamefont {Zhang}, \citenamefont {Li},\ and\
  \citenamefont {Zhu}}]{RHZ*18}%
  \BibitemOpen
  \bibfield  {author} {\bibinfo {author} {\bibfnamefont {Y.}~\bibnamefont
  {Rong}}, \bibinfo {author} {\bibfnamefont {H.~P.}\ \bibnamefont {He}},
  \bibinfo {author} {\bibfnamefont {L.}~\bibnamefont {Zhang}}, \bibinfo
  {author} {\bibfnamefont {N.}~\bibnamefont {Li}}, \ and\ \bibinfo {author}
  {\bibfnamefont {Y.~C.}\ \bibnamefont {Zhu}},\ }\href {\doibase
  https://doi.org/10.1016/j.commatsci.2018.06.023} {\bibfield  {journal}
  {\bibinfo  {journal} {Computational Materials Science}\ }\textbf {\bibinfo
  {volume} {153}},\ \bibinfo {pages} {48 } (\bibinfo {year}
  {2018})}\BibitemShut {NoStop}%
\bibitem [{\citenamefont {Zhu}\ \emph {et~al.}(2021)\citenamefont {Zhu},
  \citenamefont {Liu}, \citenamefont {Wang},\ and\ \citenamefont
  {Yang}}]{ZLWY21}%
  \BibitemOpen
  \bibfield  {author} {\bibinfo {author} {\bibfnamefont {J.}~\bibnamefont
  {Zhu}}, \bibinfo {author} {\bibfnamefont {X.}~\bibnamefont {Liu}}, \bibinfo
  {author} {\bibfnamefont {Z.}~\bibnamefont {Wang}}, \ and\ \bibinfo {author}
  {\bibfnamefont {Q.}~\bibnamefont {Yang}},\ }\href {\doibase
  10.1088/1361-651x/ac03a5} {\bibfield  {journal} {\bibinfo  {journal}
  {Modelling and Simulation in Materials Science and Engineering}\ }\textbf
  {\bibinfo {volume} {29}},\ \bibinfo {pages} {055017} (\bibinfo {year}
  {2021})}\BibitemShut {NoStop}%
\bibitem [{\citenamefont {Wang}\ \emph {et~al.}(2021)\citenamefont {Wang},
  \citenamefont {Jin}, \citenamefont {Yang}, \citenamefont {Li}, \citenamefont
  {Tang}, \citenamefont {Zong},\ and\ \citenamefont {Peng}}]{WJY*21}%
  \BibitemOpen
  \bibfield  {author} {\bibinfo {author} {\bibfnamefont {L.}~\bibnamefont
  {Wang}}, \bibinfo {author} {\bibfnamefont {J.}~\bibnamefont {Jin}}, \bibinfo
  {author} {\bibfnamefont {P.}~\bibnamefont {Yang}}, \bibinfo {author}
  {\bibfnamefont {S.}~\bibnamefont {Li}}, \bibinfo {author} {\bibfnamefont
  {S.}~\bibnamefont {Tang}}, \bibinfo {author} {\bibfnamefont {Y.}~\bibnamefont
  {Zong}}, \ and\ \bibinfo {author} {\bibfnamefont {Q.}~\bibnamefont {Peng}},\
  }\href {\doibase https://doi.org/10.1016/j.commatsci.2021.110309} {\bibfield
  {journal} {\bibinfo  {journal} {Computational Materials Science}\ }\textbf
  {\bibinfo {volume} {191}},\ \bibinfo {pages} {110309} (\bibinfo {year}
  {2021})}\BibitemShut {NoStop}%
\bibitem [{\citenamefont {Zhang}\ \emph
  {et~al.}(2020{\natexlab{b}})\citenamefont {Zhang}, \citenamefont {An},
  \citenamefont {Li}, \citenamefont {Lu}, \citenamefont {Wu},\ and\
  \citenamefont {Xia}}]{ZAL*20}%
  \BibitemOpen
  \bibfield  {author} {\bibinfo {author} {\bibfnamefont {Y.}~\bibnamefont
  {Zhang}}, \bibinfo {author} {\bibfnamefont {Q.}~\bibnamefont {An}}, \bibinfo
  {author} {\bibfnamefont {J.}~\bibnamefont {Li}}, \bibinfo {author}
  {\bibfnamefont {B.}~\bibnamefont {Lu}}, \bibinfo {author} {\bibfnamefont
  {W.}~\bibnamefont {Wu}}, \ and\ \bibinfo {author} {\bibfnamefont
  {R.}~\bibnamefont {Xia}},\ }\href {\doibase 10.1007/s00894-020-04595-y}
  {\bibfield  {journal} {\bibinfo  {journal} {Journal of Molecular Modeling}\
  }\textbf {\bibinfo {volume} {26}},\ \bibinfo {pages} {335} (\bibinfo {year}
  {2020}{\natexlab{b}})}\BibitemShut {NoStop}%
\bibitem [{\citenamefont {Kumar}(2018)}]{Kum18}%
  \BibitemOpen
  \bibfield  {author} {\bibinfo {author} {\bibfnamefont {S.}~\bibnamefont
  {Kumar}},\ }\href {\doibase
  https://doi.org/10.1016/j.matchemphys.2018.01.013} {\bibfield  {journal}
  {\bibinfo  {journal} {Materials Chemistry and Physics}\ }\textbf {\bibinfo
  {volume} {208}},\ \bibinfo {pages} {41} (\bibinfo {year} {2018})}\BibitemShut
  {NoStop}%
\bibitem [{\citenamefont {Zhang}\ \emph {et~al.}(2021)\citenamefont {Zhang},
  \citenamefont {Wang},\ and\ \citenamefont {Huang}}]{ZWH21}%
  \BibitemOpen
  \bibfield  {author} {\bibinfo {author} {\bibfnamefont {S.}~\bibnamefont
  {Zhang}}, \bibinfo {author} {\bibfnamefont {F.}~\bibnamefont {Wang}}, \ and\
  \bibinfo {author} {\bibfnamefont {P.}~\bibnamefont {Huang}},\ }\href
  {\doibase https://doi.org/10.1016/j.jmst.2021.02.013} {\bibfield  {journal}
  {\bibinfo  {journal} {Journal of Materials Science \& Technology}\ }\textbf
  {\bibinfo {volume} {87}},\ \bibinfo {pages} {176} (\bibinfo {year}
  {2021})}\BibitemShut {NoStop}%
\bibitem [{\citenamefont {Weng}\ \emph {et~al.}(2018)\citenamefont {Weng},
  \citenamefont {Ning}, \citenamefont {Fu}, \citenamefont {Hu}, \citenamefont
  {Zhao}, \citenamefont {Huang},\ and\ \citenamefont {Peng}}]{WNF*18}%
  \BibitemOpen
  \bibfield  {author} {\bibinfo {author} {\bibfnamefont {S.}~\bibnamefont
  {Weng}}, \bibinfo {author} {\bibfnamefont {H.}~\bibnamefont {Ning}}, \bibinfo
  {author} {\bibfnamefont {T.}~\bibnamefont {Fu}}, \bibinfo {author}
  {\bibfnamefont {N.}~\bibnamefont {Hu}}, \bibinfo {author} {\bibfnamefont
  {Y.}~\bibnamefont {Zhao}}, \bibinfo {author} {\bibfnamefont {C.}~\bibnamefont
  {Huang}}, \ and\ \bibinfo {author} {\bibfnamefont {X.}~\bibnamefont {Peng}},\
  }\href@noop {} {\bibfield  {journal} {\bibinfo  {journal} {Scientific
  Reports}\ }\textbf {\bibinfo {volume} {8}},\ \bibinfo {pages} {3089}
  (\bibinfo {year} {2018})}\BibitemShut {NoStop}%
\bibitem [{\citenamefont {Yousefi}\ \emph {et~al.}(2013)\citenamefont
  {Yousefi}, \citenamefont {Lin}, \citenamefont {Zheng}, \citenamefont {Shen},
  \citenamefont {Pothnis}, \citenamefont {Jia}, \citenamefont {Zussman},\ and\
  \citenamefont {Kim}}]{YLZ*13}%
  \BibitemOpen
  \bibfield  {author} {\bibinfo {author} {\bibfnamefont {N.}~\bibnamefont
  {Yousefi}}, \bibinfo {author} {\bibfnamefont {X.}~\bibnamefont {Lin}},
  \bibinfo {author} {\bibfnamefont {Q.}~\bibnamefont {Zheng}}, \bibinfo
  {author} {\bibfnamefont {X.}~\bibnamefont {Shen}}, \bibinfo {author}
  {\bibfnamefont {J.~R.}\ \bibnamefont {Pothnis}}, \bibinfo {author}
  {\bibfnamefont {J.}~\bibnamefont {Jia}}, \bibinfo {author} {\bibfnamefont
  {E.}~\bibnamefont {Zussman}}, \ and\ \bibinfo {author} {\bibfnamefont
  {J.-K.}\ \bibnamefont {Kim}},\ }\href {\doibase
  https://doi.org/10.1016/j.carbon.2013.03.034} {\bibfield  {journal} {\bibinfo
   {journal} {Carbon}\ }\textbf {\bibinfo {volume} {59}},\ \bibinfo {pages}
  {406} (\bibinfo {year} {2013})}\BibitemShut {NoStop}%
\bibitem [{\citenamefont {Chu}\ and\ \citenamefont {Jia}(2014)}]{CJ14}%
  \BibitemOpen
  \bibfield  {author} {\bibinfo {author} {\bibfnamefont {K.}~\bibnamefont
  {Chu}}\ and\ \bibinfo {author} {\bibfnamefont {C.}~\bibnamefont {Jia}},\
  }\href {\doibase https://doi.org/10.1002/pssa.201330051} {\bibfield
  {journal} {\bibinfo  {journal} {physica status solidi (a)}\ }\textbf
  {\bibinfo {volume} {211}},\ \bibinfo {pages} {184} (\bibinfo {year}
  {2014})}\BibitemShut {NoStop}%
\bibitem [{\citenamefont {Hu}\ \emph {et~al.}(2016)\citenamefont {Hu},
  \citenamefont {Tong}, \citenamefont {Lin}, \citenamefont {Nian},
  \citenamefont {Shao}, \citenamefont {Hu}, \citenamefont {Saeib},
  \citenamefont {Jin},\ and\ \citenamefont {Cheng}}]{HTL*16}%
  \BibitemOpen
  \bibfield  {author} {\bibinfo {author} {\bibfnamefont {Z.}~\bibnamefont
  {Hu}}, \bibinfo {author} {\bibfnamefont {G.}~\bibnamefont {Tong}}, \bibinfo
  {author} {\bibfnamefont {D.}~\bibnamefont {Lin}}, \bibinfo {author}
  {\bibfnamefont {Q.}~\bibnamefont {Nian}}, \bibinfo {author} {\bibfnamefont
  {J.}~\bibnamefont {Shao}}, \bibinfo {author} {\bibfnamefont {Y.}~\bibnamefont
  {Hu}}, \bibinfo {author} {\bibfnamefont {M.}~\bibnamefont {Saeib}}, \bibinfo
  {author} {\bibfnamefont {S.}~\bibnamefont {Jin}}, \ and\ \bibinfo {author}
  {\bibfnamefont {G.~J.}\ \bibnamefont {Cheng}},\ }\href {\doibase
  https://doi.org/10.1016/j.jmatprotec.2015.12.022} {\bibfield  {journal}
  {\bibinfo  {journal} {Journal of Materials Processing Technology}\ }\textbf
  {\bibinfo {volume} {231}},\ \bibinfo {pages} {143} (\bibinfo {year}
  {2016})}\BibitemShut {NoStop}%
\bibitem [{\citenamefont {Zhang}\ and\ \citenamefont {Zhan}(2016)}]{ZZ16}%
  \BibitemOpen
  \bibfield  {author} {\bibinfo {author} {\bibfnamefont {D.}~\bibnamefont
  {Zhang}}\ and\ \bibinfo {author} {\bibfnamefont {Z.}~\bibnamefont {Zhan}},\
  }\href {\doibase https://doi.org/10.1016/j.jallcom.2015.09.013} {\bibfield
  {journal} {\bibinfo  {journal} {Journal of Alloys and Compounds}\ }\textbf
  {\bibinfo {volume} {654}},\ \bibinfo {pages} {226} (\bibinfo {year}
  {2016})}\BibitemShut {NoStop}%
\bibitem [{\citenamefont {Chen}\ \emph
  {et~al.}(2016{\natexlab{a}})\citenamefont {Chen}, \citenamefont {Ying},
  \citenamefont {Wang}, \citenamefont {Du}, \citenamefont {Liu},\ and\
  \citenamefont {Huang}}]{CYW*16}%
  \BibitemOpen
  \bibfield  {author} {\bibinfo {author} {\bibfnamefont {F.}~\bibnamefont
  {Chen}}, \bibinfo {author} {\bibfnamefont {J.}~\bibnamefont {Ying}}, \bibinfo
  {author} {\bibfnamefont {Y.}~\bibnamefont {Wang}}, \bibinfo {author}
  {\bibfnamefont {S.}~\bibnamefont {Du}}, \bibinfo {author} {\bibfnamefont
  {Z.}~\bibnamefont {Liu}}, \ and\ \bibinfo {author} {\bibfnamefont
  {Q.}~\bibnamefont {Huang}},\ }\href {\doibase
  https://doi.org/10.1016/j.carbon.2015.10.023} {\bibfield  {journal} {\bibinfo
   {journal} {Carbon}\ }\textbf {\bibinfo {volume} {96}},\ \bibinfo {pages}
  {836 } (\bibinfo {year} {2016}{\natexlab{a}})}\BibitemShut {NoStop}%
\bibitem [{\citenamefont {Gao}\ \emph {et~al.}(2016)\citenamefont {Gao},
  \citenamefont {Yue}, \citenamefont {Guo}, \citenamefont {Zhang},
  \citenamefont {Lin}, \citenamefont {Yao},\ and\ \citenamefont
  {Wang}}]{GYG*16}%
  \BibitemOpen
  \bibfield  {author} {\bibinfo {author} {\bibfnamefont {X.}~\bibnamefont
  {Gao}}, \bibinfo {author} {\bibfnamefont {H.}~\bibnamefont {Yue}}, \bibinfo
  {author} {\bibfnamefont {E.}~\bibnamefont {Guo}}, \bibinfo {author}
  {\bibfnamefont {H.}~\bibnamefont {Zhang}}, \bibinfo {author} {\bibfnamefont
  {X.}~\bibnamefont {Lin}}, \bibinfo {author} {\bibfnamefont {L.}~\bibnamefont
  {Yao}}, \ and\ \bibinfo {author} {\bibfnamefont {B.}~\bibnamefont {Wang}},\
  }\href {\doibase https://doi.org/10.1016/j.powtec.2016.06.045} {\bibfield
  {journal} {\bibinfo  {journal} {Powder Technology}\ }\textbf {\bibinfo
  {volume} {301}},\ \bibinfo {pages} {601} (\bibinfo {year}
  {2016})}\BibitemShut {NoStop}%
\bibitem [{\citenamefont {Li}\ and\ \citenamefont {Xiong}(2017)}]{LX17}%
  \BibitemOpen
  \bibfield  {author} {\bibinfo {author} {\bibfnamefont {G.}~\bibnamefont
  {Li}}\ and\ \bibinfo {author} {\bibfnamefont {B.}~\bibnamefont {Xiong}},\
  }\href {\doibase https://doi.org/10.1016/j.jallcom.2016.12.147} {\bibfield
  {journal} {\bibinfo  {journal} {Journal of Alloys and Compounds}\ }\textbf
  {\bibinfo {volume} {697}},\ \bibinfo {pages} {31} (\bibinfo {year}
  {2017})}\BibitemShut {NoStop}%
\bibitem [{\citenamefont {Chen}\ \emph
  {et~al.}(2016{\natexlab{b}})\citenamefont {Chen}, \citenamefont {Zhang},
  \citenamefont {Liu}, \citenamefont {He}, \citenamefont {Shi}, \citenamefont
  {Li}, \citenamefont {Nash},\ and\ \citenamefont {Zhao}}]{CZL*16}%
  \BibitemOpen
  \bibfield  {author} {\bibinfo {author} {\bibfnamefont {Y.}~\bibnamefont
  {Chen}}, \bibinfo {author} {\bibfnamefont {X.}~\bibnamefont {Zhang}},
  \bibinfo {author} {\bibfnamefont {E.}~\bibnamefont {Liu}}, \bibinfo {author}
  {\bibfnamefont {C.}~\bibnamefont {He}}, \bibinfo {author} {\bibfnamefont
  {C.}~\bibnamefont {Shi}}, \bibinfo {author} {\bibfnamefont {J.}~\bibnamefont
  {Li}}, \bibinfo {author} {\bibfnamefont {P.}~\bibnamefont {Nash}}, \ and\
  \bibinfo {author} {\bibfnamefont {N.}~\bibnamefont {Zhao}},\ }\href {\doibase
  10.1038/srep19363} {\bibfield  {journal} {\bibinfo  {journal} {Scientific
  Reports}\ }\textbf {\bibinfo {volume} {6}},\ \bibinfo {pages} {19363}
  (\bibinfo {year} {2016}{\natexlab{b}})}\BibitemShut {NoStop}%
\bibitem [{\citenamefont {Song}\ \emph {et~al.}(2020)\citenamefont {Song},
  \citenamefont {Wang}, \citenamefont {Sun}, \citenamefont {Li}, \citenamefont
  {Sun}, \citenamefont {Fu},\ and\ \citenamefont {Pan}}]{SWS*20}%
  \BibitemOpen
  \bibfield  {author} {\bibinfo {author} {\bibfnamefont {G.}~\bibnamefont
  {Song}}, \bibinfo {author} {\bibfnamefont {Q.}~\bibnamefont {Wang}}, \bibinfo
  {author} {\bibfnamefont {L.}~\bibnamefont {Sun}}, \bibinfo {author}
  {\bibfnamefont {S.}~\bibnamefont {Li}}, \bibinfo {author} {\bibfnamefont
  {Y.}~\bibnamefont {Sun}}, \bibinfo {author} {\bibfnamefont {Q.}~\bibnamefont
  {Fu}}, \ and\ \bibinfo {author} {\bibfnamefont {C.}~\bibnamefont {Pan}},\
  }\href {\doibase https://doi.org/10.1016/j.matdes.2020.108629} {\bibfield
  {journal} {\bibinfo  {journal} {Materials \& Design}\ }\textbf {\bibinfo
  {volume} {191}},\ \bibinfo {pages} {108629} (\bibinfo {year}
  {2020})}\BibitemShut {NoStop}%
\bibitem [{\citenamefont {Varillas}\ and\ \citenamefont {Frank}(2021)}]{VF21}%
  \BibitemOpen
  \bibfield  {author} {\bibinfo {author} {\bibfnamefont {J.}~\bibnamefont
  {Varillas}}\ and\ \bibinfo {author} {\bibfnamefont {O.}~\bibnamefont
  {Frank}},\ }\href {\doibase https://doi.org/10.1016/j.carbon.2020.11.003}
  {\bibfield  {journal} {\bibinfo  {journal} {Carbon}\ }\textbf {\bibinfo
  {volume} {173}},\ \bibinfo {pages} {301} (\bibinfo {year}
  {2021})}\BibitemShut {NoStop}%
\bibitem [{\citenamefont {Wei}\ \emph {et~al.}(2013)\citenamefont {Wei},
  \citenamefont {Wang}, \citenamefont {Wu}, \citenamefont {Yang},\ and\
  \citenamefont {Dunn}}]{WWW*13}%
  \BibitemOpen
  \bibfield  {author} {\bibinfo {author} {\bibfnamefont {Y.}~\bibnamefont
  {Wei}}, \bibinfo {author} {\bibfnamefont {B.}~\bibnamefont {Wang}}, \bibinfo
  {author} {\bibfnamefont {J.}~\bibnamefont {Wu}}, \bibinfo {author}
  {\bibfnamefont {R.}~\bibnamefont {Yang}}, \ and\ \bibinfo {author}
  {\bibfnamefont {M.~L.}\ \bibnamefont {Dunn}},\ }\href {\doibase
  10.1021/nl303168w} {\bibfield  {journal} {\bibinfo  {journal} {Nano Letters}\
  }\textbf {\bibinfo {volume} {13}},\ \bibinfo {pages} {26} (\bibinfo {year}
  {2013})}\BibitemShut {NoStop}%
\bibitem [{\citenamefont {Deng}\ and\ \citenamefont {Berry}(2016)}]{DB16}%
  \BibitemOpen
  \bibfield  {author} {\bibinfo {author} {\bibfnamefont {S.}~\bibnamefont
  {Deng}}\ and\ \bibinfo {author} {\bibfnamefont {V.}~\bibnamefont {Berry}},\
  }\href {\doibase https://doi.org/10.1016/j.mattod.2015.10.002} {\bibfield
  {journal} {\bibinfo  {journal} {Materials Today}\ }\textbf {\bibinfo {volume}
  {19}},\ \bibinfo {pages} {197} (\bibinfo {year} {2016})}\BibitemShut
  {NoStop}%
\bibitem [{\citenamefont {Chen}\ \emph {et~al.}(2019)\citenamefont {Chen},
  \citenamefont {Gui}, \citenamefont {Yang}, \citenamefont {Zhu},\ and\
  \citenamefont {Tang}}]{CGY*19}%
  \BibitemOpen
  \bibfield  {author} {\bibinfo {author} {\bibfnamefont {W.}~\bibnamefont
  {Chen}}, \bibinfo {author} {\bibfnamefont {X.}~\bibnamefont {Gui}}, \bibinfo
  {author} {\bibfnamefont {L.}~\bibnamefont {Yang}}, \bibinfo {author}
  {\bibfnamefont {H.}~\bibnamefont {Zhu}}, \ and\ \bibinfo {author}
  {\bibfnamefont {Z.}~\bibnamefont {Tang}},\ }\href {\doibase
  10.1039/C8NH00112J} {\bibfield  {journal} {\bibinfo  {journal} {Nanoscale
  Horiz.}\ }\textbf {\bibinfo {volume} {4}},\ \bibinfo {pages} {291} (\bibinfo
  {year} {2019})}\BibitemShut {NoStop}%
\bibitem [{\citenamefont {Blees}\ \emph {et~al.}(2015)\citenamefont {Blees},
  \citenamefont {Barnard}, \citenamefont {Rose}, \citenamefont {Roberts},
  \citenamefont {McGill}, \citenamefont {Huang}, \citenamefont {Ruyack},
  \citenamefont {Kevek}, \citenamefont {Kobrin}, \citenamefont {Muller},\ and\
  \citenamefont {McEuen}}]{BBR*15}%
  \BibitemOpen
  \bibfield  {author} {\bibinfo {author} {\bibfnamefont {M.~K.}\ \bibnamefont
  {Blees}}, \bibinfo {author} {\bibfnamefont {A.~W.}\ \bibnamefont {Barnard}},
  \bibinfo {author} {\bibfnamefont {P.~A.}\ \bibnamefont {Rose}}, \bibinfo
  {author} {\bibfnamefont {S.~P.}\ \bibnamefont {Roberts}}, \bibinfo {author}
  {\bibfnamefont {K.~L.}\ \bibnamefont {McGill}}, \bibinfo {author}
  {\bibfnamefont {P.~Y.}\ \bibnamefont {Huang}}, \bibinfo {author}
  {\bibfnamefont {A.~R.}\ \bibnamefont {Ruyack}}, \bibinfo {author}
  {\bibfnamefont {J.~W.}\ \bibnamefont {Kevek}}, \bibinfo {author}
  {\bibfnamefont {B.}~\bibnamefont {Kobrin}}, \bibinfo {author} {\bibfnamefont
  {D.~A.}\ \bibnamefont {Muller}}, \ and\ \bibinfo {author} {\bibfnamefont
  {P.~L.}\ \bibnamefont {McEuen}},\ }\href {\doibase 10.1038/nature14588}
  {\bibfield  {journal} {\bibinfo  {journal} {Nature}\ }\textbf {\bibinfo
  {volume} {524}},\ \bibinfo {pages} {204} (\bibinfo {year}
  {2015})}\BibitemShut {NoStop}%
\bibitem [{\citenamefont {Nicholl}\ \emph {et~al.}(2015)\citenamefont
  {Nicholl}, \citenamefont {Conley}, \citenamefont {Lavrik}, \citenamefont
  {Vlassiouk}, \citenamefont {Puzyrev}, \citenamefont {Sreenivas},
  \citenamefont {Pantelides},\ and\ \citenamefont {Bolotin}}]{NCL*15}%
  \BibitemOpen
  \bibfield  {author} {\bibinfo {author} {\bibfnamefont {R.~J.~T.}\
  \bibnamefont {Nicholl}}, \bibinfo {author} {\bibfnamefont {H.~J.}\
  \bibnamefont {Conley}}, \bibinfo {author} {\bibfnamefont {N.~V.}\
  \bibnamefont {Lavrik}}, \bibinfo {author} {\bibfnamefont {I.}~\bibnamefont
  {Vlassiouk}}, \bibinfo {author} {\bibfnamefont {Y.~S.}\ \bibnamefont
  {Puzyrev}}, \bibinfo {author} {\bibfnamefont {V.~P.}\ \bibnamefont
  {Sreenivas}}, \bibinfo {author} {\bibfnamefont {S.~T.}\ \bibnamefont
  {Pantelides}}, \ and\ \bibinfo {author} {\bibfnamefont {K.~I.}\ \bibnamefont
  {Bolotin}},\ }\href {\doibase 10.1038/ncomms9789} {\bibfield  {journal}
  {\bibinfo  {journal} {Nature Communications}\ }\textbf {\bibinfo {volume}
  {6}},\ \bibinfo {pages} {8789} (\bibinfo {year} {2015})}\BibitemShut
  {NoStop}%
\bibitem [{\citenamefont {Kumar}\ \emph {et~al.}(2019)\citenamefont {Kumar},
  \citenamefont {Pattanayek},\ and\ \citenamefont {Das}}]{KPD19}%
  \BibitemOpen
  \bibfield  {author} {\bibinfo {author} {\bibfnamefont {S.}~\bibnamefont
  {Kumar}}, \bibinfo {author} {\bibfnamefont {S.~K.}\ \bibnamefont
  {Pattanayek}}, \ and\ \bibinfo {author} {\bibfnamefont {S.~K.}\ \bibnamefont
  {Das}},\ }\href {\doibase 10.1021/acs.jpcc.9b03101} {\bibfield  {journal}
  {\bibinfo  {journal} {The Journal of Physical Chemistry C}\ }\textbf
  {\bibinfo {volume} {123}},\ \bibinfo {pages} {18017} (\bibinfo {year}
  {2019})}\BibitemShut {NoStop}%
\bibitem [{\citenamefont {Mishin}\ \emph {et~al.}(1999)\citenamefont {Mishin},
  \citenamefont {Farkas}, \citenamefont {Mehl},\ and\ \citenamefont
  {Papaconstantopoulos}}]{MFMP99}%
  \BibitemOpen
  \bibfield  {author} {\bibinfo {author} {\bibfnamefont {Y.}~\bibnamefont
  {Mishin}}, \bibinfo {author} {\bibfnamefont {D.}~\bibnamefont {Farkas}},
  \bibinfo {author} {\bibfnamefont {M.~J.}\ \bibnamefont {Mehl}}, \ and\
  \bibinfo {author} {\bibfnamefont {D.~A.}\ \bibnamefont
  {Papaconstantopoulos}},\ }\href@noop {} {\bibfield  {journal} {\bibinfo
  {journal} {Phys. Rev. B}\ }\textbf {\bibinfo {volume} {59}},\ \bibinfo
  {pages} {3393} (\bibinfo {year} {1999})}\BibitemShut {NoStop}%
\bibitem [{\citenamefont {Stuart}\ \emph {et~al.}(2000)\citenamefont {Stuart},
  \citenamefont {Tutein},\ and\ \citenamefont {Harrison}}]{STH00}%
  \BibitemOpen
  \bibfield  {author} {\bibinfo {author} {\bibfnamefont {S.~J.}\ \bibnamefont
  {Stuart}}, \bibinfo {author} {\bibfnamefont {A.~B.}\ \bibnamefont {Tutein}},
  \ and\ \bibinfo {author} {\bibfnamefont {J.~A.}\ \bibnamefont {Harrison}},\
  }\href@noop {} {\bibfield  {journal} {\bibinfo  {journal} {J. Chem. Phys.}\
  }\textbf {\bibinfo {volume} {112}},\ \bibinfo {pages} {6472} (\bibinfo {year}
  {2000})}\BibitemShut {NoStop}%
\bibitem [{\citenamefont {Huang}\ \emph {et~al.}(2003)\citenamefont {Huang},
  \citenamefont {Mainardi},\ and\ \citenamefont {Balbuena}}]{HMB03}%
  \BibitemOpen
  \bibfield  {author} {\bibinfo {author} {\bibfnamefont {S.-P.}\ \bibnamefont
  {Huang}}, \bibinfo {author} {\bibfnamefont {D.~S.}\ \bibnamefont {Mainardi}},
  \ and\ \bibinfo {author} {\bibfnamefont {P.~B.}\ \bibnamefont {Balbuena}},\
  }\href {\doibase 10.1016/j.susc.2003.08.050} {\bibfield  {journal} {\bibinfo
  {journal} {Surface Science}\ }\textbf {\bibinfo {volume} {545}},\ \bibinfo
  {pages} {163} (\bibinfo {year} {2003})}\BibitemShut {NoStop}%
\bibitem [{\citenamefont {Tavazza}\ \emph {et~al.}(2015)\citenamefont
  {Tavazza}, \citenamefont {Senftle}, \citenamefont {Zou}, \citenamefont
  {Becker},\ and\ \citenamefont {van Duin}}]{TSZ*15}%
  \BibitemOpen
  \bibfield  {author} {\bibinfo {author} {\bibfnamefont {F.}~\bibnamefont
  {Tavazza}}, \bibinfo {author} {\bibfnamefont {T.~P.}\ \bibnamefont
  {Senftle}}, \bibinfo {author} {\bibfnamefont {C.}~\bibnamefont {Zou}},
  \bibinfo {author} {\bibfnamefont {C.~A.}\ \bibnamefont {Becker}}, \ and\
  \bibinfo {author} {\bibfnamefont {A.~C.~T.}\ \bibnamefont {van Duin}},\
  }\href@noop {} {\bibfield  {journal} {\bibinfo  {journal} {J. Phys. Chem. C}\
  }\textbf {\bibinfo {volume} {119}},\ \bibinfo {pages} {13580} (\bibinfo
  {year} {2015})}\BibitemShut {NoStop}%
\bibitem [{NiG()}]{NiGra_SM}%
  \BibitemOpen
  \href@noop {} {}\bibinfo {note} {See Supplementary Material for additional
  data and analyses.}\BibitemShut {Stop}%
\bibitem [{\citenamefont {Hou}\ \emph {et~al.}(2016)\citenamefont {Hou},
  \citenamefont {Dong}, \citenamefont {Tian}, \citenamefont {Liu},
  \citenamefont {Wang},\ and\ \citenamefont {Wang}}]{HDT*16}%
  \BibitemOpen
  \bibfield  {author} {\bibinfo {author} {\bibfnamefont {Z.~Y.}\ \bibnamefont
  {Hou}}, \bibinfo {author} {\bibfnamefont {K.~J.}\ \bibnamefont {Dong}},
  \bibinfo {author} {\bibfnamefont {Z.~A.}\ \bibnamefont {Tian}}, \bibinfo
  {author} {\bibfnamefont {R.~S.}\ \bibnamefont {Liu}}, \bibinfo {author}
  {\bibfnamefont {Z.}~\bibnamefont {Wang}}, \ and\ \bibinfo {author}
  {\bibfnamefont {J.~G.}\ \bibnamefont {Wang}},\ }\href {\doibase
  10.1039/C6CP02172G} {\bibfield  {journal} {\bibinfo  {journal} {Phys. Chem.
  Chem. Phys.}\ }\textbf {\bibinfo {volume} {18}},\ \bibinfo {pages} {17461}
  (\bibinfo {year} {2016})}\BibitemShut {NoStop}%
\bibitem [{\citenamefont {Kelchner}\ \emph {et~al.}(1998)\citenamefont
  {Kelchner}, \citenamefont {Plimpton},\ and\ \citenamefont
  {Hamilton}}]{KPH98}%
  \BibitemOpen
  \bibfield  {author} {\bibinfo {author} {\bibfnamefont {C.~L.}\ \bibnamefont
  {Kelchner}}, \bibinfo {author} {\bibfnamefont {S.~J.}\ \bibnamefont
  {Plimpton}}, \ and\ \bibinfo {author} {\bibfnamefont {J.~C.}\ \bibnamefont
  {Hamilton}},\ }\href@noop {} {\bibfield  {journal} {\bibinfo  {journal}
  {Phys. Rev. B}\ }\textbf {\bibinfo {volume} {58}},\ \bibinfo {pages} {11085}
  (\bibinfo {year} {1998})}\BibitemShut {NoStop}%
\bibitem [{\citenamefont {Ziegenhain}\ \emph {et~al.}(2009)\citenamefont
  {Ziegenhain}, \citenamefont {Hartmaier},\ and\ \citenamefont
  {Urbassek}}]{ZHU09}%
  \BibitemOpen
  \bibfield  {author} {\bibinfo {author} {\bibfnamefont {G.}~\bibnamefont
  {Ziegenhain}}, \bibinfo {author} {\bibfnamefont {A.}~\bibnamefont
  {Hartmaier}}, \ and\ \bibinfo {author} {\bibfnamefont {H.~M.}\ \bibnamefont
  {Urbassek}},\ }\href {\doibase 10.1016/j.jmps.2009.05.011} {\bibfield
  {journal} {\bibinfo  {journal} {J. Mech. Phys. Sol.}\ }\textbf {\bibinfo
  {volume} {57}},\ \bibinfo {pages} {1514} (\bibinfo {year}
  {2009})}\BibitemShut {NoStop}%
\bibitem [{\citenamefont {Ruestes}\ \emph {et~al.}(2017)\citenamefont
  {Ruestes}, \citenamefont {Bringa}, \citenamefont {Gao},\ and\ \citenamefont
  {Urbassek}}]{RBGU17}%
  \BibitemOpen
  \bibfield  {author} {\bibinfo {author} {\bibfnamefont {C.~J.}\ \bibnamefont
  {Ruestes}}, \bibinfo {author} {\bibfnamefont {E.~M.}\ \bibnamefont {Bringa}},
  \bibinfo {author} {\bibfnamefont {Y.}~\bibnamefont {Gao}}, \ and\ \bibinfo
  {author} {\bibfnamefont {H.~M.}\ \bibnamefont {Urbassek}},\ }in\ \href
  {\doibase 10.1002/9781119084501.ch14} {\emph {\bibinfo {booktitle} {Applied
  Nanoindentation in Advanced Materials}}},\ \bibinfo {editor} {edited by\
  \bibinfo {editor} {\bibfnamefont {A.}~\bibnamefont {Tiwari}}\ and\ \bibinfo
  {editor} {\bibfnamefont {S.}~\bibnamefont {Natarajan}}}\ (\bibinfo
  {publisher} {Wiley},\ \bibinfo {address} {Chichester, UK},\ \bibinfo {year}
  {2017})\ Chap.~\bibinfo {chapter} {14}, pp.\ \bibinfo {pages}
  {313--345}\BibitemShut {NoStop}%
\bibitem [{\citenamefont {Plimpton}(1995)}]{Pli95}%
  \BibitemOpen
  \bibfield  {author} {\bibinfo {author} {\bibfnamefont {S.}~\bibnamefont
  {Plimpton}},\ }\href@noop {} {\bibfield  {journal} {\bibinfo  {journal} {J.
  Comput. Phys.}\ }\textbf {\bibinfo {volume} {117}},\ \bibinfo {pages} {1}
  (\bibinfo {year} {1995})},\ \bibinfo {note}
  {\emph{http://lammps.sandia.gov/}}\BibitemShut {NoStop}%
\bibitem [{\citenamefont {Stukowski}(2012)}]{Stu12}%
  \BibitemOpen
  \bibfield  {author} {\bibinfo {author} {\bibfnamefont {A.}~\bibnamefont
  {Stukowski}},\ }\href@noop {} {\bibfield  {journal} {\bibinfo  {journal}
  {Model. Simul. Mater. Sci. Eng.}\ }\textbf {\bibinfo {volume} {20}},\
  \bibinfo {pages} {045021} (\bibinfo {year} {2012})}\BibitemShut {NoStop}%
\bibitem [{\citenamefont {Stukowski}\ \emph {et~al.}(2012)\citenamefont
  {Stukowski}, \citenamefont {Bulatov},\ and\ \citenamefont
  {Arsenlis}}]{SBA12}%
  \BibitemOpen
  \bibfield  {author} {\bibinfo {author} {\bibfnamefont {A.}~\bibnamefont
  {Stukowski}}, \bibinfo {author} {\bibfnamefont {V.~V.}\ \bibnamefont
  {Bulatov}}, \ and\ \bibinfo {author} {\bibfnamefont {A.}~\bibnamefont
  {Arsenlis}},\ }\href@noop {} {\bibfield  {journal} {\bibinfo  {journal}
  {Model. Simul. Mater. Sci. Eng.}\ }\textbf {\bibinfo {volume} {20}},\
  \bibinfo {pages} {085007} (\bibinfo {year} {2012})}\BibitemShut {NoStop}%
\bibitem [{\citenamefont {Stukowski}\ and\ \citenamefont
  {Arsenlis}(2012)}]{SA12}%
  \BibitemOpen
  \bibfield  {author} {\bibinfo {author} {\bibfnamefont {A.}~\bibnamefont
  {Stukowski}}\ and\ \bibinfo {author} {\bibfnamefont {A.}~\bibnamefont
  {Arsenlis}},\ }\href@noop {} {\bibfield  {journal} {\bibinfo  {journal}
  {Model. Simul. Mater. Sci. Eng.}\ }\textbf {\bibinfo {volume} {20}},\
  \bibinfo {pages} {035012} (\bibinfo {year} {2012})}\BibitemShut {NoStop}%
\bibitem [{\citenamefont {Stukowski}(2010)}]{Stu10}%
  \BibitemOpen
  \bibfield  {author} {\bibinfo {author} {\bibfnamefont {A.}~\bibnamefont
  {Stukowski}},\ }\href@noop {} {\bibfield  {journal} {\bibinfo  {journal}
  {Model. Simul. Mater. Sci. Eng.}\ }\textbf {\bibinfo {volume} {18}},\
  \bibinfo {pages} {015012} (\bibinfo {year} {2010})},\ \bibinfo {note}
  {\emph{http://www.ovito.org/}}\BibitemShut {NoStop}%
\bibitem [{\citenamefont {Misra}\ \emph {et~al.}(2005)\citenamefont {Misra},
  \citenamefont {Hirth},\ and\ \citenamefont {Hoagland}}]{MHH05}%
  \BibitemOpen
  \bibfield  {author} {\bibinfo {author} {\bibfnamefont {A.}~\bibnamefont
  {Misra}}, \bibinfo {author} {\bibfnamefont {J.~P.}\ \bibnamefont {Hirth}}, \
  and\ \bibinfo {author} {\bibfnamefont {R.~G.}\ \bibnamefont {Hoagland}},\
  }\href {\doibase 10.1016/j.actamat.2005.06.025} {\bibfield  {journal}
  {\bibinfo  {journal} {Acta Mater.}\ }\textbf {\bibinfo {volume} {53}},\
  \bibinfo {pages} {4817} (\bibinfo {year} {2005})}\BibitemShut {NoStop}%
\bibitem [{\citenamefont {Beyerlein}\ \emph {et~al.}(2015)\citenamefont
  {Beyerlein}, \citenamefont {Demkowicz}, \citenamefont {Misra},\ and\
  \citenamefont {Uberuaga}}]{BDMU15}%
  \BibitemOpen
  \bibfield  {author} {\bibinfo {author} {\bibfnamefont {I.~J.}\ \bibnamefont
  {Beyerlein}}, \bibinfo {author} {\bibfnamefont {M.~J.}\ \bibnamefont
  {Demkowicz}}, \bibinfo {author} {\bibfnamefont {A.}~\bibnamefont {Misra}}, \
  and\ \bibinfo {author} {\bibfnamefont {B.~P.}\ \bibnamefont {Uberuaga}},\
  }\href {\doibase 10.1016/j.pmatsci.2015.02.001} {\bibfield  {journal}
  {\bibinfo  {journal} {Progress in Materials Science}\ }\textbf {\bibinfo
  {volume} {74}},\ \bibinfo {pages} {125} (\bibinfo {year} {2015})}\BibitemShut
  {NoStop}%
\bibitem [{\citenamefont {Liu}\ \emph {et~al.}(2013)\citenamefont {Liu},
  \citenamefont {Yuan},\ and\ \citenamefont {Wei}}]{LYW13}%
  \BibitemOpen
  \bibfield  {author} {\bibinfo {author} {\bibfnamefont {X.}~\bibnamefont
  {Liu}}, \bibinfo {author} {\bibfnamefont {F.}~\bibnamefont {Yuan}}, \ and\
  \bibinfo {author} {\bibfnamefont {Y.}~\bibnamefont {Wei}},\ }\href {\doibase
  https://doi.org/10.1016/j.apsusc.2013.04.062} {\bibfield  {journal} {\bibinfo
   {journal} {Applied Surface Science}\ }\textbf {\bibinfo {volume} {279}},\
  \bibinfo {pages} {159 } (\bibinfo {year} {2013})}\BibitemShut {NoStop}%
\end{thebibliography}%
\newpage \clearpage



\end{document}